**Dynamical network stability analysis of multiple biological ages provides a framework for understanding the aging process**

Glen Pridham[1*], and Andrew D. Rutenberg[1†]

1. Department of Physics and Atmospheric Science, Dalhousie University, Halifax, B3H 4R2, Nova Scotia, Canada

* glen.pridham@dal.ca

†adr@dal.ca (corresponding author)

## Abstract

Widespread interest in non-destructive biomarkers of aging has led to a curse of plenty: a multitude of biological ages that each proffers a ‘true’ health-adjusted age of an individual. While each measure provides salient information on the aging process, they are each univariate, in contrast to the “hallmark” and “pillar” theories of aging which are explicitly multidimensional, multicausal and multiscale. Fortunately, multiple biological ages can be systematically combined into a multidimensional network representation. The interaction network between these biological ages permits analysis of the multidimensional effects of aging, as well as quantification of causal influences during both natural aging and, potentially, after anti-aging intervention. The behaviour of the system as a whole can then be explored using dynamical network stability analysis which identifies new, efficient biomarkers that quantify long term resilience scores on the timescale between measurements (years). We demonstrate this approach using a set of 8 biological ages from the longitudinal Swedish Adoption/Twin Study of Aging (SATSA). After extracting an interaction network between these biological ages, we observed that physiological age, a proxy for cardiometabolic health, serves as a central node in the network, implicating it as a key vulnerability for slow, age-related decline. We furthermore show that while the system as a whole is stable, there is a weakly stable direction along which recovery is slow — on the timescale of a human lifespan. This slow direction provides an aging biomarker which correlates strongly with chronological age and predicts longitudinal decline in health — suggesting that it estimates an important driver of age-related changes.

**Keywords: complexity, eigen analysis, biological age, systems biology**

## Introduction

The continued search for a biomarker of aging that quantifies the effects of natural aging and anti-aging interventions[1–3] has resulted in a proliferation of biological ages (BAs)[3], including recent epigenetic clocks[4] (such as[5–8]). Each BA estimates an individual's health-adjusted effective age, which may differ from their chronological age (CA). Each BA uses a model that converts a battery of measurements into a univariate proxy for health, either using regression on chronological age[7] or a heuristic mapping into a specific measure of health such as risk of death[5,6]. Popular BAs have been extensively validated and are often sensitive to mortality risk[9] and the effects of anti-aging interventions[2]. While it is tempting to simply pick the best BA for a particular application, or aggregate BAs using a heuristic approach[10,11], this risks missing the effects of multivariate interactions during the aging process.

Aging is putatively an interacting multivariate, multicausal process[12–15], as has borne out explicitly in computational studies[16–18]. This puts aging firmly in the purview of complexity science, where network analysis can be used to account for potentially catastrophic confounding effects due to interactions such as feedbacks between biological variables[12]. Such confounding effects could explain the conflicting results emerging from anti-aging intervention studies[2,19,20]. Learning the underlying network topology can help us to explain these confounding effects, and also to identify vulnerabilities of the biological system. For example, networks with feedback loops may be vulnerable to run-away effects[12], while bottlenecks through high-connectivity nodes may put a network at risk of complete collapse if those vital nodes are damaged[21,22]. Networks also provide a useful tool for visualizing and quantifying causal sequences: both during natural aging and, potentially, after interventions.

Fortunately, the complexity of the problem enables an elegant approach for processing and interpreting a network of interacting BAs. Dynamical stability analysis tells us that systems which are mostly stable — such as living organisms — can be understood by the way in which they respond to small perturbations. That is, we look for longitudinal disruptions to homeostasis and subsequent recovery (or lack there-of)[23]. The eigen-directions provide a spectrum of fundamental recovery rates (timescales), oriented to completely account for the complex interactions of the network. These rates describe canonical changes so that we can infer long-term behaviour by the slowest recovery rates[23,24], which determine system resilience[25]. Taking a long time to recover is indicative of weak stability and hence vulnerability to stochastic stressors, whose effects tend to pile up along slow or unstable eigen-directions. Indeed, prior work on health biomarkers showed that across 4 datasets (2 mice, 2 humans), the dominant risk direction for survival or dementia onset was always the first or second least-stable (slowest) eigen-direction[23]. Furthermore, a recent deep learning result from mice has shown the existence of an unstable latent variable[26] with properties similar to the frailty index (FI)[20,27] — notably nonlinear growth with age and sensitivity to anti-aging intervention. Since salient aging information naturally condenses into the least stable (slowest recovery) eigen-directions, these least stable eigen-directions are the best way to describe the collective aging behaviour of a network of BAs.

By using longitudinal data, we can also potentially infer causality within our network. We have already developed a generic model for homeostasis which estimates a network of interactions between biomarkers together with steady-state behaviour[23]. Although our model is trained using observational data[28], we expect that interventions that cause small perturbations will behave similarly to the random stresses which drive observational data — since these represent the random effects of interventions that individuals experience throughout their lives, such as lifestyle changes, medicine and disease. Previously, we found that including such directed cause → effect links conferred little benefit in reducing the root mean squared error (RMSE) of biomarker trajectories. Here, we revisit this question with a novel quantitative score based on predicting the correct direction of biomarker change. As a result, we obtain a model that predicts both the causal sequence of events occurring during normal aging and a best guess for how interventions will propagate.

We demonstrate that network analysis can leverage the abundance of BAs to answer fundamental questions about what causes aging and about how aging systems are likely to respond to interventions. We apply our approach to longitudinal multivariate BA data, generating a network interactome capable of capturing the coordinated effects of the BAs. By considering BAs from multiple biological scales we can surmise how the effects of aging propagate from DNA to functional decline. Once the network is estimated, we can analyse its eigen-directions to understand how the aggregate effects of multiple BAs affect the organism as time progresses. This yields dominant natural variables, which are the salient features of aging and provide canonical coordinates. We show that the least stable natural variable is an efficient choice for monitoring the aging process. We also show that the interaction network has vulnerabilities which are consistent with qualitative theories of aging, and is able to describe confounding effects of model interventions.

## Methods

### Model

We model an arbitrary dynamical system near a stable (homeostatic) point as

$$\begin{aligned} \vec{b}_{n+1} &= \vec{b}_n + \boldsymbol{W}\Delta t_{n+1}\left(\vec{b}_n - \vec{\mu}_n\right) + \vec{\epsilon}_{n+1} \\ \vec{\epsilon}_{n+1} &\sim N(0, \boldsymbol{\Sigma}|\Delta t_{n+1}|) \\ \vec{\mu}_n &\equiv \vec{\mu}_0 + \boldsymbol{\Lambda}\vec{x}_n \end{aligned} \tag{1}$$

where $\vec{b}_n$ represents an individual's set of biological ages measured at time timepoint $t_n$ and $\Delta t_{n+1} \equiv t_{n+1} - t_n$ is the measurement interval. Each individual has a different number of total measurements before leaving the study, $T_{in}$, and hence Eq. (1) applies to the $T_{in} - 1$ pairs of sequential measurements; dropout and missing data are discussed below and in the supplemental. We refer to this model as the Stochastic Finite-difference (SF) model, reflecting its relationship to the Stochastic Process model[29], which is the generalized continuous version of the SF model[23]. The model estimates: an equilibrium position, $\vec{\mu}_n$, a causal, resilience parameter, $\boldsymbol{W}$, which captures an interacting recovery network; and a noise term, $\boldsymbol{\Sigma}$ which implicitly includes additional effects not in the model — such as non-linear effects and fast dynamical changes. The model is only sensitive to changes which occur slower than the timescale set by $\Delta t_{n+1}$, which for this study is approximately 3 years. This means that short term changes such as due to a flu infection appear in the noise term, $\boldsymbol{\Sigma}$. $\boldsymbol{W}$ captures only long term "resilience", i.e. longitudinal correlations, over the course of years — such as age-related decline. The equilibrium position, $\vec{\mu}_n$, is allowed to vary linearly with respect to a set of covariates for each individual, $\vec{x}_n$, through $\boldsymbol{\Lambda}$.

The diagonal elements of $\boldsymbol{W}$ permit recovery towards $\vec{\mu}_n$, whereas the off-diagonals couple values across BAs,

$$\frac{E\left(b_{jn+1} - b_{jn}\right)}{E(\Delta t_{n+1})} = W_{jj}E\left(b_{jn} - \mu_{jn}\right) + \sum_{k \neq j} W_{jk}E(b_{kn} - \mu_{kn}), \tag{2}$$

where $E(x)$ represents the expectation value of $x$. (In deriving Eq. (2) we assume that the current BA values, $b_{jn}$, are negligibly correlated with the follow-up time, $\Delta t_{n+1}$.) This means that if — through intervention or natural aging — the values of some of the BAs change, $b_k$, then we will see a change in each different $b_j$ via $W_{jk}$. The diagonal elements $W_{jj}$ are the marginal recovery rates of $b_j$ ignoring all other BAs whereas the off-diagonal elements $W_{jk}$ allow interactions between different BAs — hastening or ameliorating their decline. Hence changes to one BA can propagate into the other BAs, allowing for a central driver of either age-related dysfunction or anti-aging treatment. The system only stops when each $b_j$ simultaneously reaches its equilibrium position, $\mu_{jn}$. As we will see, the estimated equilibrium positions can only be reached well beyond the lifespan of normal humans and hence the system drifts indefinitely. The interactions, $\boldsymbol{W}$, can be simplified by diagonalization using the eigen-decomposition to yield a set of composite natural aging variables $z_k$ which satisfy

$$\frac{E(z_{kn+1} - z_{kn})}{E(\Delta t_{n+1})} = \lambda_k E(z_{kn} - \tilde{\mu}_{kn}) \tag{3}$$

for $\vec{z} \equiv \boldsymbol{P}^{-1}\vec{b}$ and $\vec{\tilde{\mu}} \equiv \boldsymbol{P}^{-1}\vec{\mu}$, where $\lambda_k$ is the associated eigenvalue. We index the $z_k$ by their sorted eigenvalue strength, such that $z_1$ has the greatest (closest to $+\infty$) eigenvalue $\lambda_1$. (For simplicity, we drop the tilde notation for the remainder of the paper.) Observe that the natural aging variables, $z_k$, do not interact: they either increase, decrease or stay the same, depending on the value of $Re(\lambda_k)$ as can be seen by iterating Eq. (3). ($Im(\lambda_k) \neq 0$ contributes oscillations, but we focus exclusively on the real parts of eigenvalues in the present study.) The timescale over which these changes occur are set by $|Re(\lambda_k)|^{-1} \equiv |\lambda_k|^{-1}$, i.e. the absolute value of the real part of $\lambda_k$ ($\lambda_k$ has units of $\text{years}^{-1}$). The inverse timescale, $\lambda_k$, determines how quickly the average individual reaches a steady-state. Since the $z_k$ are independent via Eq. (3), events or interventions which modulate only $E(z_k)$ will not affect any other $E(z_j)$. The payoff of this approach is two-fold: aging information gets compressed into a few specific variables, and the interconnected system behaviour is greatly simplified.

The $z_k$ are able to drive the observed BAs through the mapping $\boldsymbol{P}\vec{z} = \vec{b}$, which will spread out the effects across several BAs since $\boldsymbol{P}$ is often dense[23]. In general, the slowest $z_k$ (greatest $\lambda_k$) have the slowest recovery ($\lambda_k > 0$ never recover); previously we observed that the key $z_k$ driving changes in $\vec{b}$ are always among the slowest[23] (for health biomarkers).

The eigen-decomposition also lets us decompose the network, represented as a matrix of weights $\boldsymbol{W}$, into a sum of sub (eigen)-networks (matrices),

$$\boldsymbol{W} = \sum_i \lambda_i P_{\cdot i} \otimes P_{i\cdot}^{-1} \tag{4}$$

where $P_{\cdot i}$ is both the $i$th column of $\boldsymbol{P}$ and the $i$th eigenvector, and $\vec{x} \otimes \vec{x} \equiv \vec{x}\vec{x}^T$ defines the outer product, $\otimes$. Each eigenvalue, $\lambda_i$, has associated with it a sub-network, $P_{\cdot i} \otimes P_{i\cdot}^{-1}$. The network is the sum of all sub-networks, weighted by their associated eigenvalues (e.g. Supplemental Figure S2). This permits us to visually analyse the effective network for each $z_k$ using their associated eigenvalue-eigenvector pair, $\lambda_k$ and $P_{\cdot k} \otimes P_{k\cdot}^{-1}$.

We estimated $\boldsymbol{\Lambda}$, $\boldsymbol{W}$ and $\boldsymbol{\Sigma}$ using linear regression as described in the supplemental. We also iteratively impute the expected model mean for all missed measurements, as described in Missing Data.

### Data

We use publicly available longitudinal data from Li *et al.*[9] Their data are derived from the Swedish Adoption/Twin Study of Aging (SATSA), and include: age, sex and 9 BAs. The population included N=845 individuals (342 males), average age at entry: $63.6 \pm 0.3$ years (standard deviation: 8.6, min: 44.9, max: 88.0). Individuals were regularly measured with median $\Delta t = 3$ years (inter-quartile range: 2.3-3.4 years) and a median number of 4 measurements per person (inter-quartile range: 3-7 measurements, max: 9). Survival data were not included in the dataset, and patients were instead labelled as dropouts after their last measurement.

The biological ages used in the present analysis are summarized in Table 1. We considered 9 BAs from 4 biological scales: genetic, epigenetic, system, and entire organism. BAs were harmonized to the same scale (~years) as follows: Telomere was multiplied by 69.29 and then we added 20.72 to match the standard deviation and mean of CA. Similarly, Cognition was multiplied by 0.9440 then we added 21.34. We model the dynamics of the 8 BAs ("predictors") and hold out the FI as a longitudinal health outcome, since it is a good predictor of risk of adverse outcome such as morbidity and mortality[18,30] (the FI is also non-Gaussian[27], in contrast to our model assumptions). We fit a more general network including the FI and CA in the supplemental.

### Statistics and Data Handling

All analysis and statistics were performed using R version 4.1.1[31]. Errors were estimated by bootstrapping using 100 resamples, unless otherwise specified. All statistical tests are z-tests, unless otherwise specified. All error bars are standard errors, unless specified otherwise. Fitting and simulating functions, as well as fitted parameter values, are available on GitHub at https://github.com/GlenPr/stochastic_finite-difference_model.

Table 1: Biological Age Summary[9]

| Biological age | Risk direction[(a)] | Scale | Input | Output/Pooling |
|---|---|---|---|---|
| Frailty index (FI)[(b)] | up | organism | health deficits[(c)] | mean |
| Functional aging index (FAI) | up | organism | sensory, grip, pulmonary and gait[(d)] | standardize then average |
| Cognition | **down** | system (brain) | cognitive testing | PC1 |
| Physiological Age (PhysioAge) | up | system (cardiometabolic) | biomarkers and physical exam[(e)] | PCA then Klemera-Doubal[7] |
| GrimAge | up | epigenetic[(f)] | CpGs | mortality risk[(g)] |
| PhenoAge | up | epigenetic[(h)] | CpGs | mortality risk[(i)] |
| Hannum | up | epigenetic | CpGs | CA[(j)] |
| Horvath | up | epigenetic | CpGs | CA[(j)] |
| Telomere | **down** | genetic | $\frac{\text{Telomere length}}{\text{standard deviation}}$ | — |

(a) Direction of change with increasing chronological age.
(b) Reserved as an outcome measure of individual health.
(c) Score from 0 (none) to 1 (full): disability, disease, and self-reported ill-health.
(d) Self-reported hearing/vision, grip strength, lung strength and gait speed.
(e) Male: body mass index, waist-to-height ratio, weight, systolic blood pressure, diastolic blood pressure, hemoglobin, serum glucose (log), and apolipoprotein B. Female: hip circumference, waist circumference systolic blood pressure, serum glucose (log), and triglycerides (log).
(f) Trained to emulate smoking pack years and plasma proteins: adrenomedullin, beta-2-microglobulim, cystatin C, GDF-15, leptin, PAI-1, and tissue inhibitor metalloproteinases 1.[6]
(g) Linearly transform mortality risk to match mean/standard deviation of chronological age.[6]
(h) Trained to predict time-to-death which includes a proportional hazard from: albumin, creatinine, serum glucose, C-reactive protein, lymphocytes (%), mean red cell volume, red cell distribution width, alkaline phosphatase, and white blood cell count.[5]
(i) Invert 10 year multivariate mortality risk (Gompertz + proportional hazard with 9 covariates).[5]
(j) CA: chronological age.

## Preprocessing

Before fitting, we transformed the BAs at each timepoint using principal component analysis (PCA) where the transformation was learned from the first timepoint (except for the diagonal model). The transformation is isomorphic (information-preserving)[23] so we estimated model parameters in PC-space then mapped them into BA-space. The number of PCs to use was selected by minimizing the 632-corrected root mean squared error (RMSE), which was 8 (max/information preserving). 632-correction uses a linear mixture of 63.2% out-of-sample test error and 36.8% in-sample training error[23]. We used PCA because selecting fewer PCs than BAs can avoid collinearity — as was done in the supplemental when the FI and CA were included in the network. *A priori*, Telomere was initially batch adjusted using linear regression[32], we observed that Telomere was normally distributed but included a few extreme outliers (right tail). Since these could be artifacts of the batch adjustment, we excluded all outliers with $p < 10^{-5}$ ($9/6006 \ll 1\%$ of entries).

## Model Selection

For initial model selection, we minimize the RMSE and mean absolute error (MAE). We used 632-corrected error values, since these have minimal bias for our model[23]. For ties, we maximize the area under the receiver operator characteristic curve (AUC)[33] of the 8 pooled BAs worsening in the next timestep, which is the probability that the prediction will correctly rank individuals who will see an increase in BA as higher than those who will not[34]. We select between a fully flexible $\boldsymbol{W}$ ("FullW") and three simplified versions: the null model (with $\boldsymbol{W} = \boldsymbol{0}$), diagonal in BA-space ("DiagW"), or diagonal in principal component–space ("SymW", which has symmetric $\boldsymbol{W}$).

## Missing Data

Data were missing due to missed measurements and dropout at an overall rate of 76%. We imputed the missed measurements and considered the effect of imputing dropped patients in the supplemental — the latter made no visible difference to the final results. Excluding dropout, the majority of ("predictor") BA values were missing (53%), which broke down as the following missingness: 20% (PhysioAge), 23% (Cognition), 27% (FAI), 60% (Telomere), 74% (Horvath), 74% (Hannum), 74% (PhenoAge), and 74% (GrimAge). The FI was missing in 20% of cases.

Missing data were initially imputed by carrying forward the last measurement, then reversed and carried backwards, then we imputed any remaining missingness using the mean of a multivariate Gaussian independently for each timepoint. This initial imputation was replaced at each fit iteration (×5) by the mean model prediction (expectation-maximization). See supplemental for full details.

Failure to impute could lead to biased conclusions[35] since most missingness in clinical studies is due in part to poor health[36]: here we observed that individuals missing all epigenetic BA measurements were significantly older ($p = 10^{-10}$, Wilcox test). Imputed values for these BAs were higher than observed, ostensibly accounting for this effect. The relatively high missingness makes imputation quality important. Imputation quality was visually assessed as good, with realistic dispersion, trajectories and age-dependence (Supplemental Figures S3 and S4). The available case analysis had much lower significance levels, but captured most of the coarse grained features of the imputed analysis (Supplemental Figures S7 and S8). Multiple imputation may give a better estimate of true effect sizes since it accounts for imputation uncertainty, see Supplemental Figure S8; qualitative results were identical to our primary imputation result. We consider only the singly imputed analysis in the main text. All outcome measures consider only observed values.

**Results**

We compared several model variants, notably: the full model according to Eq. (1) (FullW), the diagonal model Eq. (3) either in the PCA basis (SymW) or with the raw BA variables (DiagW), and the null model with $\boldsymbol{W} = \boldsymbol{0}$. The diagonal elements parameterize self-recovery from perturbations while the off-diagonal elements parameterize interactions between the variables. The symmetrical $\boldsymbol{W}$ has only undirected links (bidirectional interactions)**.** The RMSE and MAE were worse both for the null model and for the non-interacting model (DiagW) — but did not discriminate the asymmetric $\boldsymbol{W}$ (FullW) from the symmetric (SymW). This indicates that interactions were present and important for prediction ($\boldsymbol{W} \neq \boldsymbol{0}$ and not diagonal). To break the tie between FullW and SymW, we picked the one which best predicted the direction of change in BA at the next timestep (BA went up in 60% of measurements and down in 40%). We found that FullW performed better at predicting this worsening, having both lower mean absolute error at 68% confidence and higher AUC at $p = 0.1$ using the Delong test[33] (combined: $0.04 \leq p \leq 0.1$). Our final model (FullW) predicted future values with accuracy $R^2_{train} \approx R^2_{test} = 0.65 \pm 0.01$, $\text{RMSE}_{632} = 5.75 \pm 0.09$ years, and worsening AUC of $0.764 \pm 0.005$.

The interaction network, $\boldsymbol{W}$, estimated from the data is presented in Figure 1. $\boldsymbol{W}$ indicates that PhysioAge is the central node and the primary driver of changes over time (the strongest total outgoing links), with GrimAge as an important secondary, high-connectivity node. Observe that there is a positive feedback loop between the highest-connected nodes, PhysioAge → GrimAge → PhysioAge. Explicit inclusion of CA and the FI into the network does not appreciably change the connectivity of these nodes (Supplemental Figure S22); nor does the choice of imputation strategy (Supplemental Figure S7). Note that in the supplemental we confirm that the FI is not connected to Hannum, Horvath, PhenoAge or GrimAge, as was reported elsewhere using an unrelated statistical model[37]. Returning to Figure 1, the high connectivity of PhysioAge would allow it to very quickly propagate dysfunction via Eq. (2).

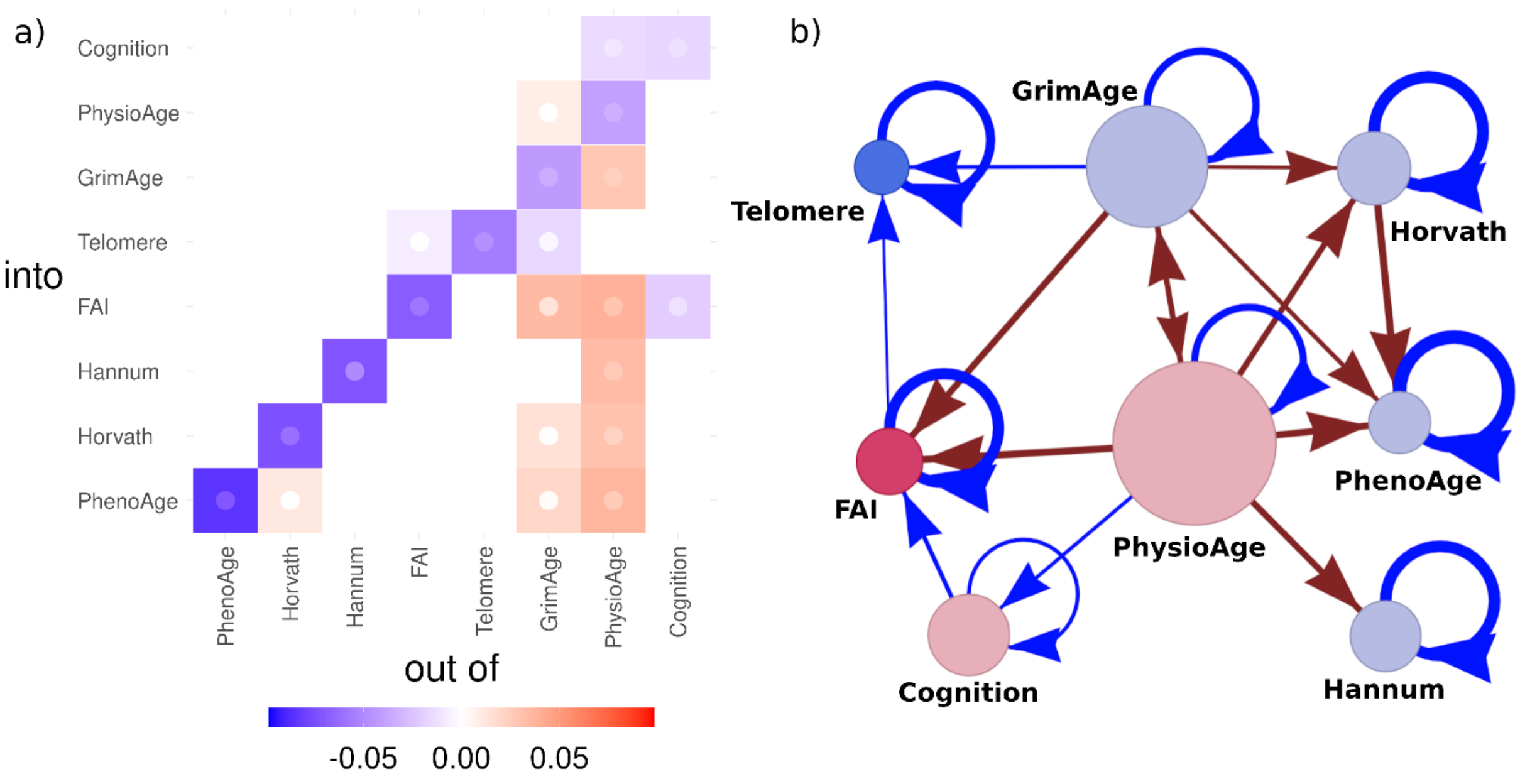

Figure 1: **Network interactome**. Both representations: matrix **(a)** and network **(b)** are equivalent. PhysioAge is the dominant node with the strongest connections, directly driving almost all other BAs (but not Telomere $p = 0.3$). GrimAge has weaker connections but also has many outgoing connections. All links are significant at $p < 0.05$. **a) Network weight matrix, $\boldsymbol{W}$**. Our model estimates each interaction parameter in this matrix. Inner point is limit of 95% CI closest to 0: point is most visible for the least significant tiles. Non-significant tiles are whited-out (p > 0.05). The elements leaving PhysioAge are typically larger than those leaving GrimAge, and have higher statistical significance (Supplemental Figure S8). Both have weak diagonal recovery $W_{jj}$. Matrix is rank-ordered by diagonal recovery strength. **b) Network representation.** Networks encode conditional dependence structures: variables are conditionally dependent if and only if there is a link directly connecting them. For example, GrimAge and Cognition are conditionally independent, since they only interact via an intermediary (PhysioAge). Node size, $n_k = \sqrt{\sum_{j \neq k} W_{jk}^2}$ (outgoing strength). Node colour indicates biological scale (see Table 1).

Drift of the BAs with age is the result of the pursuit of equilibrium, $\mu_j$. $\mu_{0j}$ ranged from ($-120 \pm 100$ to $35 \pm 24$ years) for BAs which decrease with age (Cognition and Telomere) and (127 $\pm$ 38 to 190 $\pm$ 82 years) for the increasing BAs (remaining). (Sex-effects were small, $\leq$ 10 years; Supplemental Table S1.) In all cases the equilibrium position is far outside of the age distribution of the population, causing them to drift coherently with age in their respective risk direction (Supplemental Figure S14). In the z-picture this effect is concentrated into $z_1$ and $z_2$ which drift the most, and $z_3$ which saturates around age 90 — the remaining $z_k$ quickly equilibrated and stopped changing with age (Supplemental Figure S13). This is an indication that age-related changes are concentrated into the slowest natural variables, $z_1$, $z_2$ and $z_3$, and primarily into $z_1$.

The eigenvalues of $\boldsymbol{W}$ determine system stability, so our focus is on the greatest eigenvalues which therefore recover slowest (i.e. those closest to zero since they are negative; mean-stability is determined by Eq. (3)). The eigenvalues are presented in Figure 2a. Both $z_1$ and $z_2$ (green triangles) are notably slower than the slowest diagonal elements $W_{11}$ and $W_{22}$ (orange points). The associated timescales are $|\lambda_1|^{-1} = 127 \pm 53$ years and $|\lambda_2|^{-1} = 44 \pm 8$ years. Observe that both timescales are on the order of a typical human lifespan and are significantly longer than the remaining lifespan of the population, which were all older adults (baseline ages 45-88). The timescales, $|\lambda_k|^{-1}$ determine how quickly the $z_k$ converge to the steady-state (Eq. (3)). This means that the long-time behaviour of the system will depend increasingly on $z_1$ and $z_2$, which will dominate the mapping into the BAs via $\vec{b} = \boldsymbol{P}\vec{z}$. In Figure 2b we visualize the $\lambda_1$-eigenvector using Eq. (4). The $\lambda_1$-eigenvector is centered on a fully-outgoing-connected PhysioAge with feedbacks between GrimAge and Cognition. This means that $z_1$ represents the collective action of these 3 BAs driving changes in all 8 BAs. While these 3 BAs already have the slowest marginal recoveries, $W_{jj}$, the collective action of $z_1$ is even slower due to interactions between the BAs

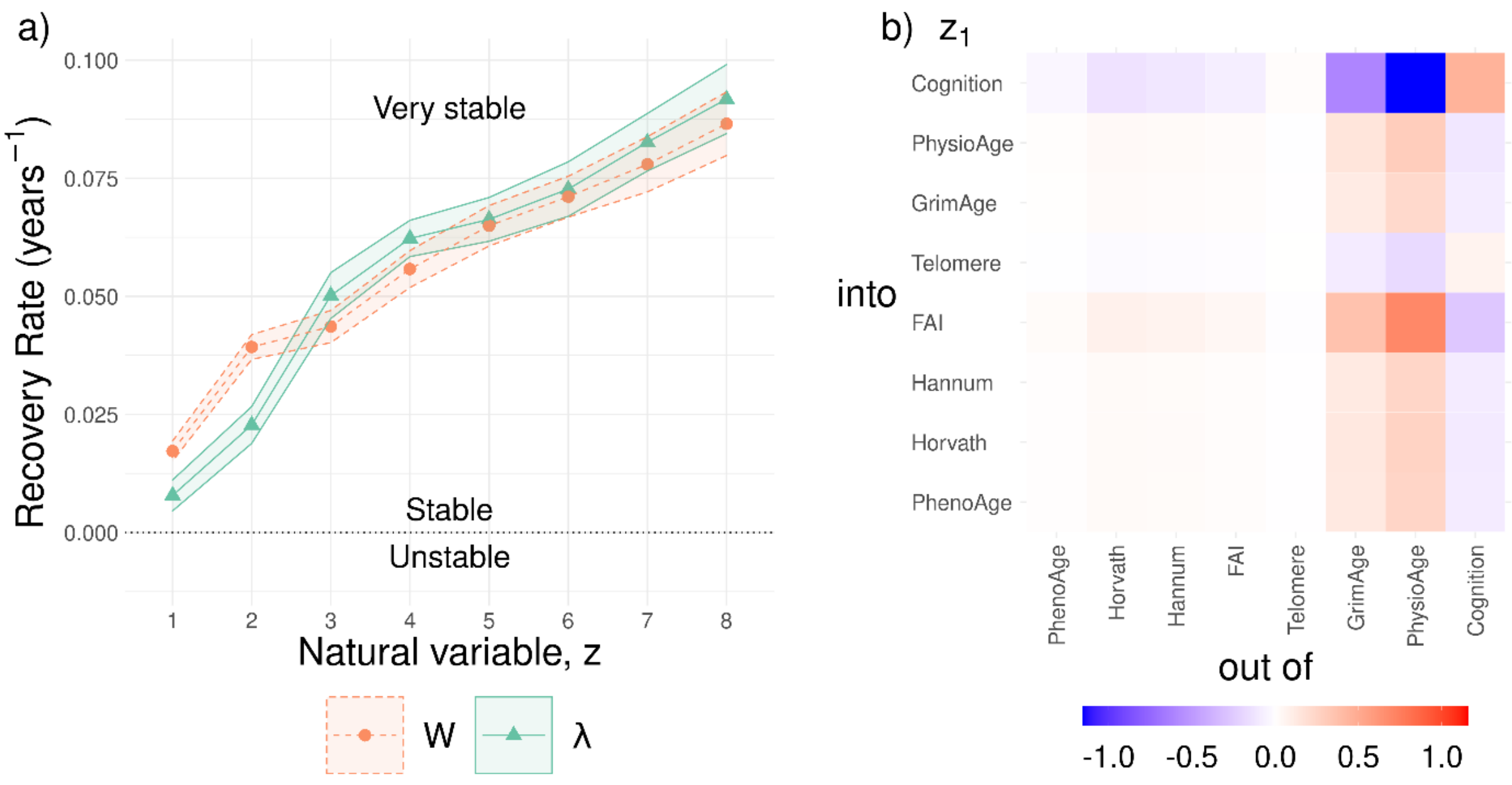

Figure 2: **Natural variables of $\boldsymbol{W}$**. Natural variables do not interact, allowing us to analyse their stability. We observed a very weak stability **(a)**, indicative of a slow recovery rate, $\lambda_1 \approx 0$. The associated eigenvector is visualized in **(b)**, comparing to Figure 1a we see that the slowest eigenvector captures the dense outgoing connections from GrimAge and PhysioAge, including feedback loops. **a) Network stability (resilience).** Eigenvalues, $\lambda_k$, determine recovery rate, $-\lambda_k$. While the network diagonal ($W_{jj}$) indicates that some biomarkers recover slowly, the network as a whole recovers even slower along $\lambda_1$ and $\lambda_2$. (Eigenvalue rank is used to index the $z_k$.) **b) Eigen-network of $z_1$.** Associated with each eigenvalue is an eigenvector. The matrix of eigenvectors, $\boldsymbol{P}$, is used to generate the natural variables as linear combinations of BAs ($\vec{z} = \boldsymbol{P}^{-1}\vec{b}$). The slowest recovering/least stable direction, $z_1$, is predominantly PhysioAge, Cognition and GrimAge, all connected into the remaining BAs. Plotted is $P_{1\cdot} \otimes P_{1\cdot}^{-1}$, where $P_{1\cdot}$ is the first eigenvector (Eq. (4)). Note the role of well-connected BAs with feedback loops: the $z_1$ eigen-network has links both above and below the diagonal.

This bottlenecking of aging information into $z_1$ and, to an extent $z_2$, is easily confirmed by looking at the correlation matrix, Figure 3a. $z_1$ is strongly correlated with almost every BA (weakly with Telomere), and always in the same risk direction. $z_2$ shares these correlations except for Cognition, suggesting that splitting between $z_1$ and $z_2$ is primarily due to differences in cognitive aging rate. Both also had the strongest correlations with chronological age (CA) and the FI of any $z_k$. Multivariate ANOVA confirmed that the real part of $z_1$ was the dominant predictor of the FI (59% of the explained variance). Correlations with the FI were concentrated into the lowest $z_k$, which is clearly demonstrated in the multiply imputed correlations Supplemental Figure S9 (which accounted for imputation error). In the supplemental we demonstrate that $z_1$ and $z_2$ are the furthest from equilibrium, which causes them to drift for the entire human lifespan leading to the observed correlation with CA e.g. Figure 3b. Altogether, it appears that the observed age-related changes, including health, are concentrated into the least stable dimensions, particularly $z_1$.

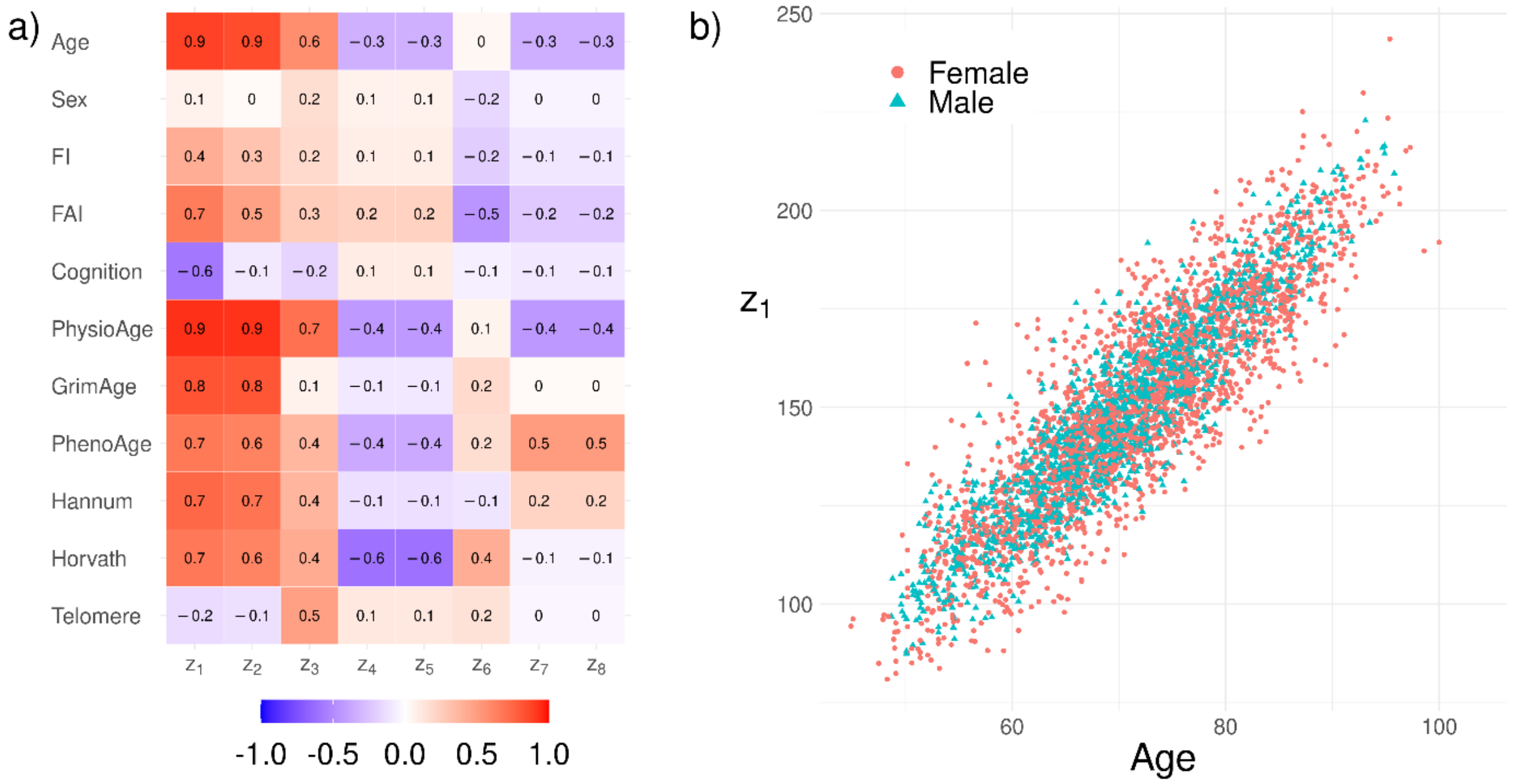

Figure 3: **Natural variable correlates. a) Spearman correlations.** $z_1$ is strongly correlated with each other BA, CA and the FI (weakly with Telomere). **b) $z_1$ is strongly correlated with CA**. As such, $z_1$ is capturing essential aging information, including CA and individual health (as estimated by the FI). (Imputed values are included only for the $z_k$.)

Our model, Eq. (1), encodes causal dependence of the current timepoint on the previous timepoint. Hence we can simulate the dynamics after a hypothetical intervention using Eq. (1) and the estimated model parameters. We operationalize interventions as an instantaneous rejuvenation of a targeted BA at a specific CA, in a manner which emulates the switching of mortality risk immediately due to an anti-aging intervention[38]. We simulated matched case and control populations for various interventions, each population contains 50000 males and 50000 females and starts at age 60; initial values were sampled from the fully-observed BAs of sex-matched individuals in the age range 55-65. We simulated using Eq. (1) with timesteps of 1 year. We include a simple model for the FI as a function of the BAs and CA, $R^2 = 0.30$, to demonstrate how to track the expected change in health as a function of the BAs (details in supplemental).

Having observed the central role of PhysioAge in Figure 1, we simulate the impact of a beneficial intervention administered at age 70 which instantly rejuvenates PhysioAge by 10 years (Figure 4). The intervention causes complex, delayed effects in the other BAs, including an adverse effect: a small, transient telomere shortening (all other BAs improved). This is due to the intervention effect propagating through the network. For example, Telomere worsens (shortens) for about 5 years post-intervention then recovers and ultimately improves after about 10 years post-intervention. Observe that, in contrast to the BAs, the relative FI *continuously improves* with time post-intervention. This is due to the unstable nature of the FI: which grows exponentially with age[27] due to compounding (propagating) secondary damage. Conversely, if we simulate an adverse event, say of disease, which increases PhysioAge by 10 years then we see the same effects with the sign flipped (Supplemental Figure S15). The long-term consequences of the adverse event continue to worsen the relative health (FI) of the case versus control even after the disease. This is consistent with results from a computational network model of disease which show the long-term FI-effect is due to secondary, compound ("propagated") damage[39].

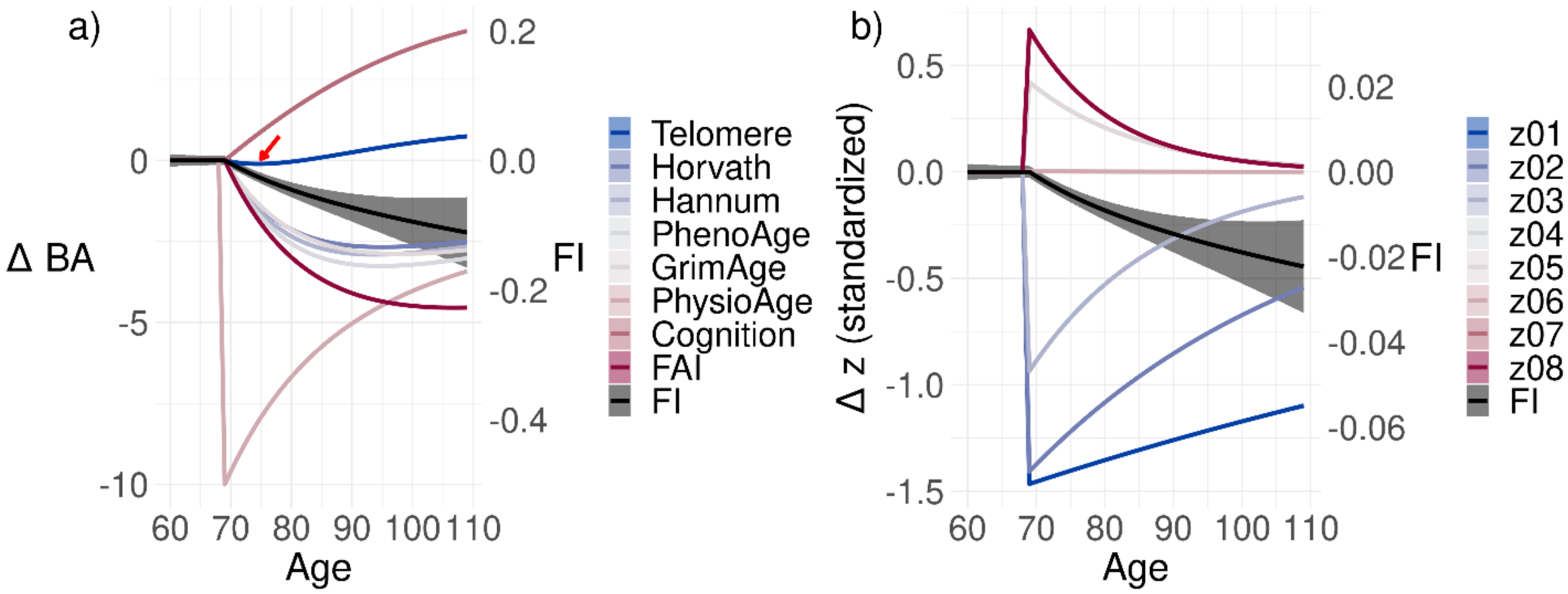

Figure 4: **Simulated intervention on PhysioAge**. We simulated a hypothetical intervention at age 70 which instantly rejuvenates PhysioAge by 10 years. **a) BA-picture**. $\Delta BA \equiv BA_{case} - BA_{control}$. We see an immediate rejuvenation of PhysioAge at age 70 due to the intervention, whereas the remaining BAs have complex delayed effects. For example, Telomere temporarily worsens (arrow) then recovers and ultimately rejuvenates. **b) z-picture**. $\Delta z \equiv z_{case} - z_{control}$. When working with natural variables, the same intervention effect is immediately spread across all $z_k$ which are connected to PhysioAge through $\vec{b} = \boldsymbol{P}\vec{z}$ ($\boldsymbol{P}$ is dense). Each $z_k$ simultaneously responds then the $z_k$ with fast recovery times quickly revert back leaving only the slow recovery times, $z_1$ and $z_2$. For the FI, the relative improvement gets better over time since compound (propagated) damage is avoided by the rejuvenation (the FI is unstable). The FI has its own y-scale as indicated. $z_k$ have been standardized for convenient comparison (zero-mean, unit-

variance). The sign of $z_k$ is arbitrary due to idiosyncrasies of the eigen-decomposition, but could be aligned using their correlation with age or health. Case and control have been perfectly matched for age, sex and stochastic effects. See supplemental for other simulated interventions. Band is standard error (often smaller than line width).

It is much easier to understand the effects of the intervention on PhysioAge using the natural variables, $z_k$, as shown in Figure 4b. All natural variables, $z_k$, connected to PhysioAge immediately improve upon the rejuvenation, then the $z_k$ return to normal on the timescale set by $|\lambda_k|^{-1}$, which means that after 20 years all of the fast $z_k$ have mostly recovered. This explains why we should primarily concern ourselves with the greatest eigenvalues, $\lambda_1$ and $\lambda_2$, since their respective natural variables are the only ones with lasting, long-term impacts — good or bad.

Intervening directly on the natural variables, $z_k$, gives a greatly simplified picture (Supplemental Figure S16). Because the $z_k$ don't interact with each other, the intervention pinpoints one $z_k$ that immediately improves whereas the other $z_j$ are unaffected. Depending on the stability of the intervened $z_k$, the effect of the intervention is either gradually lost with age (stable), persists indefinitely (marginal stability/slow dimensions $|\lambda_k| \sim 1$ lifespan$^{-1}$), or improves with increasing age (unstable). In the present study, interventions which improve $z_1$ are most desirable since they have the strongest relationship with health (Figure 3) and persist for a typical human lifespan. See the supplemental for other simulated interventions.

**Discussion**

The ongoing proliferation of new BAs (biological ages) presents two opportunities for dynamical network analysis: (1) BAs can be used to generate network interactomes to better understand how age-related changes are naturally orchestrated and for comparison to theory, and (2) there is an increasing need for a robust method of aggregating multiple BAs. Here we address both opportunities. Using a dynamical model with minimal assumptions we are able to estimate an interaction network from a collection of BAs. We can also apply eigen-analysis to the network such that we are able to generate dynamically-independent aggregate BAs — natural aging variables. The slow recovery and age-related drift of these variables reflects their underlying importance in quantifying the aging process. We propose that such natural variables are the natural language to communicate the aging process, in the same manner that spectral signal analysis has come to dominate many quantitative disciplines.

The estimated interaction network, $\boldsymbol{W}$, encodes conditional dependencies across BAs of the current timestate from the previous timestate. This permits causal predictions. For example, $W_{ij} > 1$ indicates that if $b_j$ is lower than $\mu_j$ at timepoint $t_n$ then it will push down the $i$th variable ($b_i$) by time $t_{n+1}$ (Eq. (2)). We observed that the $\mu_j$ are large enough such that the $b_j$ are always pushing each other and are autonomously drifting towards worse health. This drift fills the same role as a "mallostatic" drift with age[23], $\mu(t)$, but without explicit inclusion of time. High outgoing-degree nodes, such as PhysioAge and GrimAge, play a key role since they push the other BAs, such that changes to PhysioAge or GrimAge naturally propagate into the other, downstream BAs (e.g. rejuvenation). These relationships are learned from observational changes, rather than interventional[28]. However, we know that individuals experience many interventions throughout their lives due to medical interventions, lifestyle changes and stressors of living, such as disease. This suggests that our observation, $\boldsymbol{W}$, should be consistent with perturbative (small) interventions. In support, we found that we were better able to predict worsening of BAs by including causal relationships.

Using multiple BAs we were able to estimate an informative network of interactions which can enhance our knowledge of the aging process. In the present study, we used BAs of varying biological scales ranging from genetic (Telomere) to whole organism (FAI). We observed that PhysioAge, representing primarily cardiometabolic system changes, was the central node with outgoing arrows directly affecting all other BAs (the Telomere link was not significant). This means that changes to PhysioAge will propagate to the other BAs, Eq. (2). This permits PhysioAge to drive the other BAs. We observed a weaker, but similarly well-connected effect emanating from GrimAge including feedbacks with PhysioAge. GrimAge had a strong association with mortality for this dataset (similar to the FI)[9], and may represent damage. This implies that cardiometabolic, system-level dysfunction is essential to age-related changes, complemented by genetic/epigenetic damage with feedbacks between the two scales. Both PhysioAge and GrimAge had the strongest Spearman correlations with CA at 0.90 and 0.79, respectively (Hannum was third with 0.73). This supports the interpretation that age-related changes emerge first in those two variables then propagate outwards, causing the correlation with CA to drop as the information gradually attenuates through the network connections. The key natural variable, $z_1$, has a strong association with PhysioAge with important contributions from GrimAge, suggesting that well-connected nodes with feedbacks caused the weak stability, which appears to be primarily related to metabolic functioning.

These observations are consistent with theory. Systems biology informs us that metabolism is a vulnerable point due to its bottleneck ("bow-tie") through glucose[12,22]; its large number of outputs also make it well suited for propagating dysfunction — just as we observed with PhysioAge. Notably, metabolism is considered 1 of 7 "pillar" causes of aging, another is macromolecular damage — which is ostensibly captured by GrimAge[14]. The hallmark theory of aging specifically includes epigenetic changes and genomic instability as 2 of 12 "hallmarks" — which GrimAge may be sensitive to — but is considerably less specific towards metabolic changes, grouping cardiometabolic changes into generic changes to intercellular communication[15]. This could be an indication that the hallmark theory lacks specificity, although our suite of BAs may be similarly limited. Our suite of available BAs constrains our model to effective dynamics[26], which may differ from the oracle truth — although we can use CA as a catch-all for unaccounted degrees of freedom such as unmeasured BAs (supplemental). Fortunately, BAs are becoming increasingly specific in response to demands for high-dimensional representations of aging[2], such as biological system-specific ages[11]. As new BAs emerge, we can continue to use our approach to refine our understanding of the causal relationships underlying aging.

While the network topology is informative, the dynamical behaviour of the system as a whole is obfuscated by its complexity. Eigen-analysis identifies weakly stable, slow-recovery eigenvalues and associated eigenvectors, which can be used to generate aggregate biomarkers — i.e. natural variables. The means of natural variables do not interact, giving them simple and intuitive dynamical behaviour over time: stable natural variables simply decay towards $\mu_j$ on a timescale of eigenvalue$^{-1}$ according to Eq. (3). Previously we observed that across 4 datasets (2 mouse, 2 human) the dominant natural variable for predicting survival or dementia onset was either the first or second slowest eigenvalue, which were always stable but near 0 — specifically they took an entire lifespan to recover[23]. Ostensibly, these natural variables capture irreversible changes, such as damage. Across studies, we consistently observe that the greatest eigenvalues (slowest; most positive) are associated with the salient features of aging. In the present study these were $z_1$ and $z_2$ which had the strongest correlations with CA and were associated with health. In contrast to *ad hoc* approaches to finding the best aggregate BA[10,11], the natural variables are essential features of the dynamical system as a whole. As such, they do not interact and they become increasingly dominant with age, as they store information from stochastic events, making them dominate all of the BAs. The implication is that aging becomes increasingly simple and dependent on these essential natural variables with age (effectively leading to reduced dimensionality[18]). This is a consequence of the mapping $\vec{b} = \boldsymbol{P}\vec{z}$ which leverages redundancies in the BA representation to reduce the dimensionality. This means that aging may be lower dimension that popular theories suggest[14,15]. What's more, the dimensionality of aging may be perturbative in the sense that the first dimension provides the most important information and the subsequent dimensions provide less and less, consistent with previous computational studies[17,18]. In general, we see that the long-time dynamical behaviour is dominated by the slowest network eigenvalues, representing the directions of slowest recovery and, by implication, the lowest resilience[25].

There is significant interest in using BAs to quantify the effects of anti-aging interventions[1,2], with some going so far as to *define* rejuvenation by its prolonged effect on biological age[2]. Our simulated interventions highlight the utility of dynamical stability (eigen) analysis in disambiguating network dynamics and identifying optimal intervention targets. Using our estimated model parameters we were able to simulate a hypothetical intervention at age 70 that instantly rejuvenates PhysioAge by 10 years. Improvement to a BA post-intervention has been observed in a number of experiments with epigenetic BAs[2] and other BAs, such as the FI[20]. A consistent problem that emerges from these experiments is that health may improve along one dimension at the expense of another, such as increased tumorigenesis post-rejuvenation[2], reduced visual acuity following anti-aging treatment via metformin[20], and increased frailty following mTORC2 disruption[19] — which is inhibited by rapamycin treatment. Such pleiotropic effects are mediated through some biological interaction network — albeit typically an unknown one.

We observed a similar pleiotropic effect in our simulated intervention, wherein 7/8 of the BAs showed immediate or delayed rejuvenation but worsening was observed in Telomere (shortening) during a 10 year long transient effect. The general issue is that the interaction network obscures the effects of interventions on a single BA. Such a seemingly simple intervention perturbs the network, which then adjusts all of the BAs through its connections — leading to delayed and unexpected downstream effects. This problem can be avoided by working in the natural variables, $z_k$. Because the mapping is linear and invertible, we can easily transform between the BA and $z$–pictures as needed. When working in the natural variables the intervention is greatly simplified: all stable $z_j$ revert to the control with a timescale $|\lambda_j|^{-1}$, and all unstable $z_k$ improve continuously post-intervention with timescale $|\lambda_k|^{-1}$. This simplifies the problem since we immediately know that $|\lambda_j|^{-1}$ is the length of time that a stable $z_j$ will differ from control, so we only need to monitor the key natural variables which exhibit high risk and low recovery: $\lambda_k \gtrsim 0$ — the remaining $z_j$ will quickly forget the perturbation. This requires only that we determine the relevance of each $z_k$ to health, which can be done prior to an interventional study (and may naturally compress into the lowest $z_k$). Identifying unstable and weakly stable natural variables, and the interventions that modulate them should be a fruitful topic of future research.

An unstable natural variable would be particularly important, although we have now failed to observe an instability in 5 datasets using the SF model, using either BAs or health biomarkers[23]. An instability would lead to super-linear growth, such as are observed in the frailty index (FI)[27], dynamical FI[26], and specific plasma proteins[8]. Since natural variables drive observed variables via $\vec{b} = \boldsymbol{P}\vec{z}$, unstable natural variable(s) could be of prime importance since all stable natural variables should eventually equilibrate such that all observed age-related changes are driven by the unstable natural variable(s). Furthermore, amelioration of an unstable natural variable would result in continuous life-long improvements such as we saw with the FI in our simulated interventions. However, the vast majority of health biomarkers change linearly with age[40], and BAs are typically designed to track CA in units of time which, by definition, increases linearly with age. The choice of units used to quantify aging may therefore play a role in determining what super-linear growth means, and therefore stability. A second issue is the effect of a population-level picture which can mask unstable sub-populations, such as those experiencing or transitioning into chronic disease. This presents an opportunity for more powerful statistical models able to capture such individual effects. Note that a slow instability is indistinguishable from linear drift until advanced ages — where data are sparse. Our current perspective is that organisms live most of their lives in a stable regime of approximately linear decline until a tipping point is reached and non-linear collapse ensues, quickly leading to organism death or chronic disease.

While linear drift is reasonable population-level behaviour for the BAs and $z_k$, a tipping point leading to super-linear behaviour is necessary to make sense of terminal decline. Terminal decline occurs in biomarkers of death wherein immediately prior to death they become much worse, driving up the hazard and dropping the survival probability towards 0 over a short period of time. This effect can be seen empirically, for example, in the FI[41], cognition[42] and gait[42] which show a dramatic change in slope occurring around 3-4 years prior to death. Two dynamical phases are needed to capture this effect, such as those elucidated by the saturating-repair model of aging[43,44]. In this model, age-related decline begins as a stable, approximately linear system until repair processes “saturate” and a new super-linear phase begins. This model unifies ideas of critical behaviour[26,45,46] with damage and repair[47,48], which have been increasingly used within the aging modelling literature. Our prior results from mice and humans supports the important predictions made by the saturating-repair model, including “mallostasis”[23]: the correlation between mortality hazard and linear drift rate[43]. The linear phase is characterised by a linear increase in the homeostatic steady-state and is ostensibly driven by asymmetric transitions such as epigenetic methylation and accumulation of disease[49]. We hypothesize that the linear phase serves to push individuals towards tolerance thresholds (tipping points) upon which a super-linear phase ensues, quickly leading to death or disease e.g. due to the saturation of repair processes.

It is important to understand that our model is of slow, linear dynamics in $\boldsymbol{W}$ at timescales slower than the interval between measurements (years). Faster dynamics are pushed into the noise term, $\boldsymbol{\Sigma}$. The effects of the fast dynamics are to cause biomarkers to rapidly change in values from day to day, whereas the slow dynamics estimated by $\boldsymbol{W}$ are long term changes over the course of human-equivalent years. The resilience score we estimate using $\boldsymbol{W}$ represents resilience on the timescale of years, which (we believe) is a good timescale to assess aging. Nevertheless, this is in contrast to typical measurements of biological resilience which are on the short timescale of weeks and shows a clear age-dependence[45]. In contrast, we observed no clear age-dependence for the slow-resilience assessed by our $\boldsymbol{W}$-analysis, noting the large error bars (supplemental). A clear age-dependent drop in the eigenvalues of $\boldsymbol{W}$ could be an indication of saturating repair, but the effect is indirect and bounded at 0,[43,44] which may make it difficult to observe. The correct interpretation of $\boldsymbol{W}$-resilience is currently unclear. $\boldsymbol{W}$ is capturing the long-term decline due to aging through the longitudinal correlations it causes in the biological ages. The smallest eigenvalues have the longest memories and hence they are the natural place for information regarding long term and irreversible changes to build up, making the associated eigenvectors excellent predictors of age-related health.

Our relatively modest predictive performance of worsening (AUC $0.764 \pm 0.005$ ), and explained FI variance of approximately 30%, suggests that we have only captured some of the age-related changes. Our model achieved a prediction error on the order of approximately 6 years for BA progression after an average of 3 years of natural aging, representing 65% of the variance. Model error encapsulates the net effect of four major sources: missing values, unobserved variables, stochasticity, and model misspecification. The missingness was particularly high in the present study, especially for the epigenetic BAs, which were more likely to be missing for older individuals (thus non-random), which could lead to bias — even with good imputation[35]. This missingness reduces data quality and quantity, which may explain why we did not achieve statistical significance when comparing only the AUC of FullW vs SymW ($p = 0.1$). Second, we relied on only 8 BAs to completely predict future health including only sex as a covariate. This limits predictive power and our ability to detect causal relationships, since we cannot identify causal connections from unobserved variables[50]. While it is impossible to capture all information, we would hope to find a saturation 'elbow' at some larger number of BAs. Third, in addition to intrinsic stochasticity, there is substantial stochasticity owing to non-lab conditions (individual variability) and measurement noise, putting severe limits of predictability. Finally, our model is a local approximation near stable homeostasis and does not capture sudden changes, such as may occur due to the emergence of an underlying chronic condition. While the model could incorporate sudden changes via $\mu_n$, this is only possible when they are specified.

There is a trend towards increasingly specific BAs which are compatible with non-redundant, multivariate representations[11,18]. Our approach complements these representations, since it provides both causal structure and generates salient aggregate features. Our approach is not limited to BAs, it works for all continuous-valued, longitudinal biomarkers. For example, emerging 'omics data, such as dynamical changes to proteiomics[8] could be a natural target for dynamical network stability analysis. We have previously applied the approach to physiological biomarkers including blood tests, body weight, blood pressure and other generic health biomarkers[23]. While correlation analysis is the *de facto* standard, it has a severe shortcoming in that it estimates unconditional relationships which therefore cannot represent a true network (they do not satisfy the standard axioms of graph theory[51] and hence their interpretation is ambiguous at best). A network link indicates a conditional relationship given all variables in the network, permitting easy and intuitive interpretation. Our approach is very general, and we suggest analysts consider using it any time they apply correlation analysis to longitudinal data. We think that the greatly enhanced interpretability is worth the modest additional computational burden. For more quantitative researchers, our linear model could easily be replaced with a more complex model, such as a deep neural network[17], and the analysis of independent natural variables could remain the same. The key natural variables we observe may also be useful measures for secondary analysis, such as looking for early warning signs for the onset of chronic disease[52].

Our central hypothesis is that BAs provide the raw information needed to generate interaction networks, which can then be analysed as a whole using dynamical stability (eigen) analysis. Our work highlights the utility of approaches borrowed from complexity science and systems biology. Our results are consistent with aging having several of the key features of complex systems: networks, motifs and feedbacks, which we show play an important role in understanding age-related changes. This is direct evidence of the importance of complex and systems-level thinking[12] for furthering our understanding of aging. In summation, a simple, interpretable model of the dynamics can be leveraged to estimate the essential effects of aging, and infer the effects of perturbative interventions. We demonstrate that analysing our fitted data gives results consistent with known theory, making it a potential path forward to operationalizing and testing various qualitative theories of aging. Far from being a curse of plenty, the proliferation of established, new and increasingly-specific BAs may be the key to quantifying and understanding the complex multidimensional changes which characterize the aging process.

## Conflict of Interest

None.

## Funding

This work was supported by the Natural Sciences and Engineering Research Council of Canada (NSERC; grant number RGPIN-2019-05888) to ADR.

## Data Availability

Data are publicly available from Li *et al.*[9] Our fitting and simulating functions using R are available online at https://github.com/GlenPr/stochastic_finite-difference_model. Exact model parameters are provided in CSV files on GitHub.

## Acknowledgements

GP built the model and analysed the data. ADR and GP conceived the project. All authors reviewed the manuscript.

**Supplemental information for “Dynamical network stability analysis of multiple biological ages provides a framework for understanding the aging process”**

Glen Pridham[1*], and Andrew D. Rutenberg[1†]

1. Department of Physics and Atmospheric Science, Dalhousie University, Halifax, B3H 4R2, Nova Scotia, Canada

* glen.pridham@dal.ca

†adr@dal.ca (corresponding author)

## Summary

This supplemental contains additional information that supports and expands our analysis in the main text. We start with two important utility results: a model for the FI which can be used to transform the FI into a normal random variable, and an extended explanation of the eigen-decomposition of a network matrix into a sum of sub-network matrices. Next we provide additional information on the materials and methods, in particular the biological ages (BAs) used, the imputation process, and important results for estimating our model. The missing data handling section includes reiteration of our key results using alternative methods including multiple imputation and available case. We then provide additional results that supplement the main text. Finally, we end with sensitivity analysis where we consider sex and age–stratified fits, and then the inclusion of additional variables in our network analysis (the FI and chronological age, CA).

## Frailty Index (FI) Model

The FI is the average number of health deficits an individual has, which ranges from 0 (perfect health) to 1 (all deficit). We wish to treat the FI like the other BAs but this constrained range causes issues with our model, which assumes normally distributed errors. In this section we seek a transformation of the FI that makes it behave more like a BA: in particular with improved normality. Simultaneously, we seek a model to track FI dynamics during our simulated interventions. In this section we demonstrate that a simple phenomenological model for the FI also naturally yields a transformation that makes the FI approximately normally distributed, solving both issues.

We start by looking for a model for the FI in order to track the simulated FI trajectory. The canonical model is

$$f = f_0 \mathrm{e}^{\alpha t} \tag{S1}$$

where $f$ is the FI and $t$ is the age in years ($\alpha \approx 0.035 \pm 0.001$ years$^{-1}$ and $\ln(f_0) \approx -4.16 \pm 0.001$ are fit parameters[27]). This implies that the FI satisfies the differential equation

$$\frac{df}{dt} = \alpha f \tag{S2}$$

but unfortunately, the model suffers from the "zero state" problem[27]: if $f = 0$ at any time then it will stay at 0 indefinitely. This is unappealing both conceptually and practically: since $f = 0$ is measured in the present study (32/3162 of FI measurements were 0). The simplest solution to this issue is to introduce a constant damage rate $\gamma$ such that

$$\frac{df}{dt} = \alpha f + \gamma. \tag{S3}$$

$\gamma$ can be thought of as the rate of external damage, whereas $\alpha$ captures the net effect of compound (propagated) damage. This is physically plausible and avoids the zero state problem.

The solution of Eq. (S3) is easily confirmed to be

$$f = f_0 \mathrm{e}^{\alpha t} - \frac{\gamma}{\alpha}. \tag{S4}$$

What about covariates? We have several measures of biological age, $\vec{b}$, which, we assume can be treated as additional information that complements the chronological age, $\alpha t \rightarrow \vec{\beta}^T \vec{b}$ (where $\vec{b}$ includes CA (chronological age)). This yields a log-linear model,

$$\ln(f + \frac{\gamma}{\alpha}) = \vec{\beta}^T \vec{b} + \ln(f_0). \tag{S5}$$

We selected CA and FAI as $\vec{b}$. We hypothesized then verified that under this transformation the stochastic component of the FI can be approximated as a normal distribution yielding the final model

$$\ln(f + \frac{\gamma}{\alpha}) = \vec{\beta}^T \vec{b} + \ln(f_0) + \xi, \tag{S6}$$

where $\xi \sim N(0, \sigma')$ is a normal random variable (the model residual). It is known empirically that the FI has a strongly skewed distribution[27] bounded at 0 which will be strictly enforced by the exponential inverse transformation. The transformed FI is plotted against age in Figure S1, along with the model residual, $\xi$. The interpretation of this transformation is that it converts the FI, $f$, into a linear function of age with normally-distributed error term i.e. a biological age on its native scale. We expect normally-distributed errors for the population's biological age since the individuals are all over age 40 and hence should have been exposed to many stochastic events that could age or rejuvenate them leading to a normal distribution by the central limit theorem.

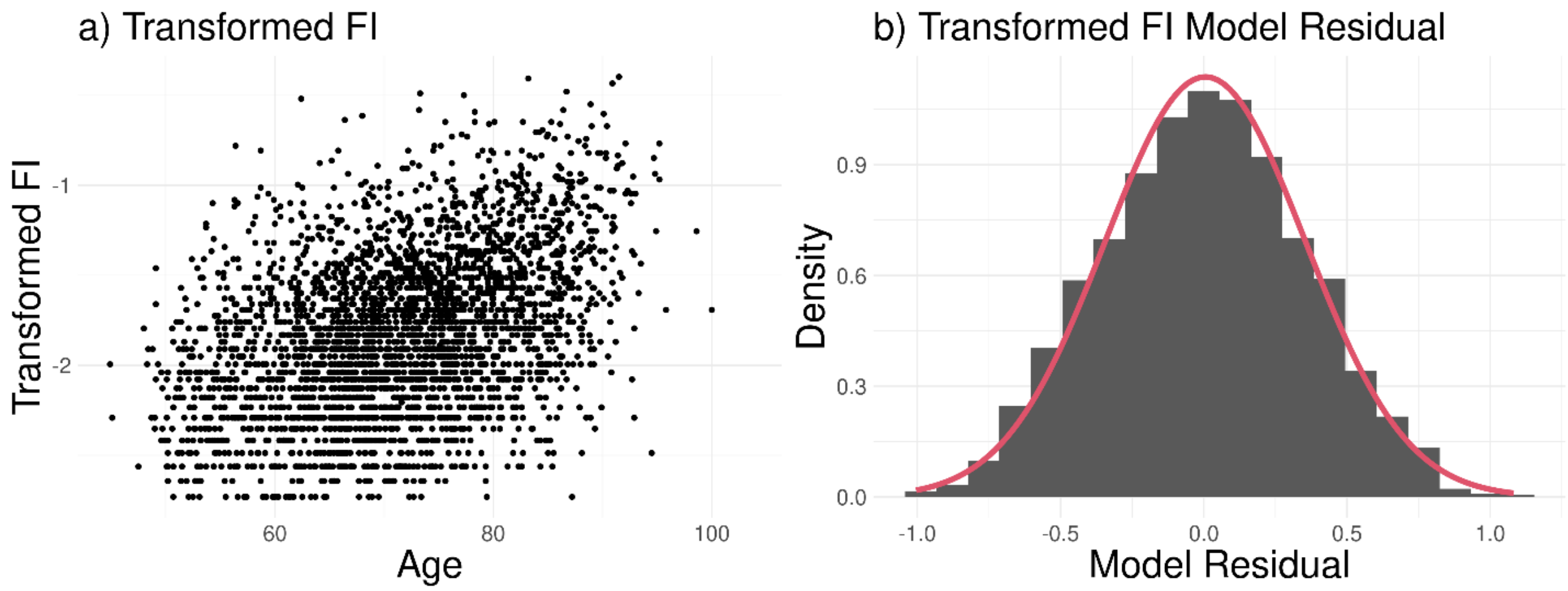

Figure S1: **Transformed FI.** The FI was transformed by $ln(f + 0.065)$ to improve linearity **(a)** and produce an unbiased error term **(b)**. **a)** the transformed FI behaves like a BA having direct proportionality to age, it can be scaled to yield an FI-equivalent BA — which we did in Sensitivity Analysis. **b)** model residual, $\xi$. The FI is skewed due to being bound at 0, the transformation removes that skewness and instead we have a normally-distributed error term (red line).

Observe that if we can estimate $\gamma/\alpha$ then we can estimate the remaining terms in Eq. (S6) using linear regression. We iteratively fit a linear model using linear regression and picked the parameter $\gamma/\alpha$ that minimized the absolute residual skewness. This yielded the estimate $\gamma/\alpha = 0.065$.

Minimizing the skewness is a technical assumption to get a transformed $f$ which is symmetrical, this is needed because the loss functions we use are symmetrical. If the transformed $f$ was asymmetrical then the fitted models would ignore extreme values, instead fitting to the bulk of the distribution. This effect is seen with the untransformed $f$, for which the linearized models we use ignore individuals near $f = 0$ which leads the models to greatly over-estimate $f$ for healthy individuals (not shown).

In summary, the FI, $f$, is linearized by the invertible transformation $\ln(f + 0.065)$, allowing us to use linear models in our analysis. Our FI model in the simulations uses the transformation purely as a technical pre-processing step to improve the regression fit. Our network model in Sensitivity Analysis however treats the transformed FI as another BA to include in the model. Readers should be aware that this network model predicts the transformed FI, not the true FI. The transformed FI is an unscaled biological age with slope equal to the damage propagation rate ($\alpha$). Before fitting to the network in Sensitivity Analysis we scaled to units of age by matching the mean and standard deviation of CA. The use of the FI as a biological age is discussed by Mitnitski and Rockwood[27].

**Eigen-decomposition**

Central to our analysis is the eigen-decomposition, which decomposes the network weight matrix, $\boldsymbol{W}$, into a set of eigenvalues and respective eigenvectors. The eigen-decomposition of $\boldsymbol{W}$ can be written as

$$\boldsymbol{W} = \sum_{i=1}^{p} \lambda_i P_{\cdot i} \otimes P_{i\cdot}^{-1} \qquad \text{(S7)}$$

where $\boldsymbol{W}$ has dimensions $p \times p$, $\boldsymbol{P}^{-1}\boldsymbol{W}\boldsymbol{P}$ is diagonal and $P_{\cdot i}$ is the $i$th column (and eigenvector); note $\vec{x} \otimes \vec{x} = \vec{x}\vec{x}^T$. Hence $\boldsymbol{W}$ is a linear sum of the sub-networks, $P_{\cdot i} \otimes P_{i\cdot}^{-1}$, as shown in Figure S2. Note that the sub-networks, $P_{\cdot i} \otimes P_{i\cdot}^{-1}$, are constrained, rank-1 matrices — this is the proximal cause of their blocky appearance (see e.g. Figure S2a, $P_{\cdot 1} \otimes P_{1\cdot}^{-1}$). The weakly stable eigenvalues have blocky, high-connectivity nodes and feedbacks in their sub-networks because for a sub-network to be strong enough to out-weigh the diagonal in the eigen-decomposition, Eq. (S7), it needs to have a large block of uniform values — which are precisely high-connectivity nodes with feedbacks.

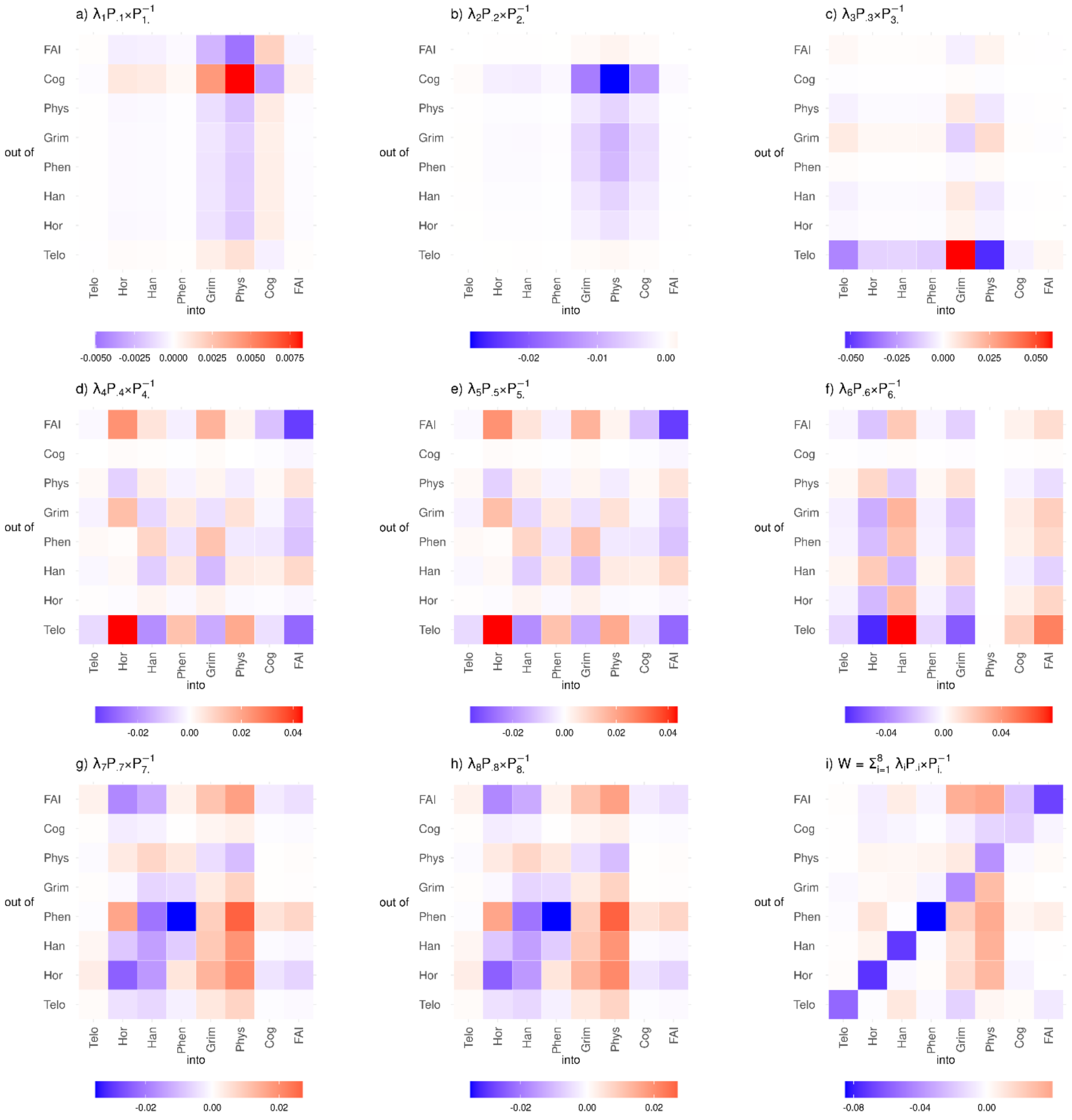

Figure S2: **Network eigen-decomposition**. The network weight matrix, $\boldsymbol{W}$, can be represented as a sum of eigenvalues and sub-networks (proportional to the eigenvectors). The sub-networks ranked by eigenvalue from 1 to 8 are given in **a-h**. **i)** is equivalently the sum of the first 8 and $\boldsymbol{W}$, in accordance with Eq. (S7). Observe the high-connectivity blocks in **a)** and **b)** are associated with low stability. Having links above and below diagonal indicates feedbacks (bi-directional links). Only real components are shown; **d/e** and **g/h** are repeated eigenvectors which differ only in their imaginary components. Causal links flow from "out of" to "into", where the sign of the causal effect is indicated by the colour (red is positive effect, while blue is negative).

## Materials and Methods

Here we provide some additional details on the dataset and data handling, starting with a description of the BAs used.

Telomere length is leukocyte T/S-ratio (qPCR assay product over reference)[32]. Batch adjustments of telomere length were made by the original authors using linear regression[32], which justified our choice to exclude extreme outliers, as described in the main text. Horvath and Hannum both capture age-associated epigenetic changes via penalized regression, irrespective of their relationship to health[9]. GrimAge and PhenoAge are more refined since they capture effective survival risk age. GrimAge uses an intermediate step such that only epigenetic changes associated with smoker pack-years or specific proteins are used, the latter are associated with a variety of conditions including inflammation, cardiometabolic dysfunction and cellular functioning[6]. PhenoAge, in contrast, uses epigenetic changes to predict individualized mortality risk using a Gompertz model of survival with proportional hazards[5]. PhysioAge estimates a latent (unobservable) variable which captures mutual age-dependence between the variables and chronological age, each biomarker and chronological age are modelled as a linear function of the latent variable[7]. The variables used are overwhelmingly cardiometabolic, hence PhysioAge likely represents age-related changes to the cardiometabolic system. Cognition is a general cognitive ability score: the first principal component from a battery of cognitive tests[53]. FAI is based on self-reported sensory ability (hearing and seeing), and measured: lung strength, grip strength and gait speed[54]. The FI included 42 measures of health, including self-reported health and sensory, diseases, symptoms, activities of daily living (ADLs) and instrumental ADLs[55] (ADLs are needed to maintain an individual's basic needs; instrumental ADLs are more complex activities needed to live independently[56]).

## Missing Data

The majority of ("predictor") BA values were missing (53%), which broke down as the following missingness: 20% (PhysioAge), 23% (Cognition), 27% (FAI), 60% (Telomere), 74% (Horvath), 74% (Hannum), 74% (PhenoAge), and 74% (GrimAge). The FI was missing in 20% of cases.

Prior to fitting, missing data were initially imputed as follows:

1. Carry forward value from previous timepoint to next timepoint,
2. if still missing, carry backwards value from next timepoint to previous timepoint (missingness after steps 1 and 2: 1% (PhysioAge and Cognition), 3% (FAI), 10% (Telomere), 41% (Horvath, Hannum, PhenoAge and GrimAge)),
3. if still missing, impute the time-independent conditional Gaussian mean[57] using other measurements at that same timepoint (missingness after steps 1-3: 0%). (The BAs in this dataset are known to be strongly correlated through their mutual CA-dependence[9].)

After the initial imputation, all of the imputed values were re-imputed using the model mean at each iteration of fitting algorithm (×5).[23] Specifically, we first carried back the current value of the second timepoint estimate into the first timepoint (to avoid inverting $\mathbf{W}$, which has small determinant), then we forward imputed all timepoints past the first using the model mean.

We assessed imputation quality visually both at the population level and for randomly sampled individuals. The population level imputation looked good (Figure S3): the age-dependence and dispersion are similar between both the imputed and observed values. The epigenetic ages are observably high in the mean for most ages, but this is expected because we know that the sub-population who were measured were younger ($p = 10^{-10}$, Wilcox test). We expect the true missing values to be higher than the observed, consistent with what was imputed.

The individual level imputation also looked reasonable, for example Figure S4 shows 10 randomly sampled individuals. We see that the individual trajectories follow the overall population trend while generally interpolating smoothly between observed timepoints (keeping in mind that the biomarkers have multivariate dependencies). We deemed the overall imputation quality was good.

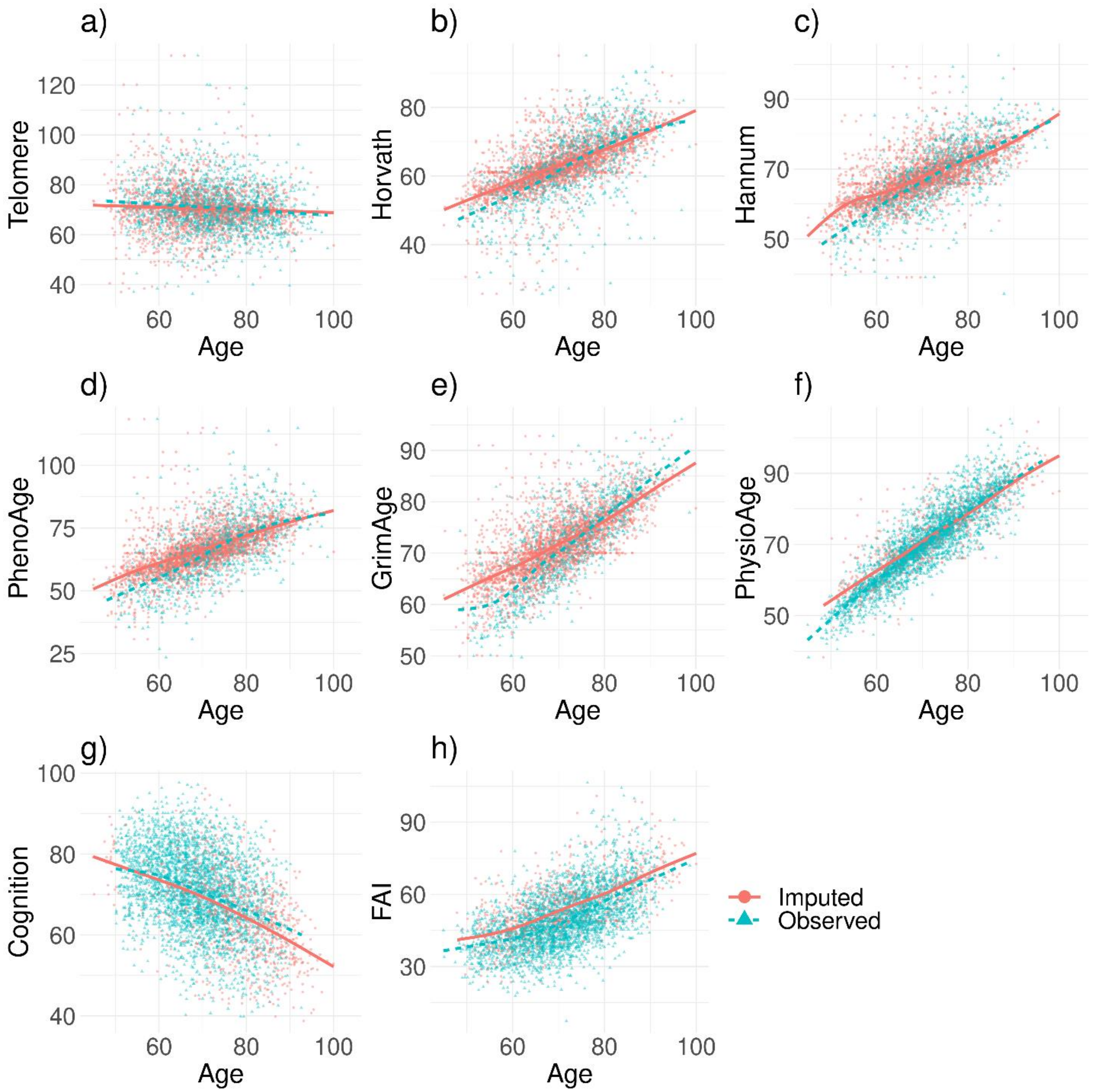

Figure S3: **Imputation quality — age-distribution**. Imputed values (red points) have nearly identical distributions to the observed values (blue triangles): similar dispersion, mean, and age-dependence. The epigenetic ages do, however, show that the majority of imputed values are larger than the observed, this is expected since the individuals missing those values tended to be older (Horvath, Hannum, PhenoAge and GrimAge). Altogether these observations indicated a good imputation. Lines are best fits from a cubic spline additive model using the MGCV package with default degrees of freedom[58].

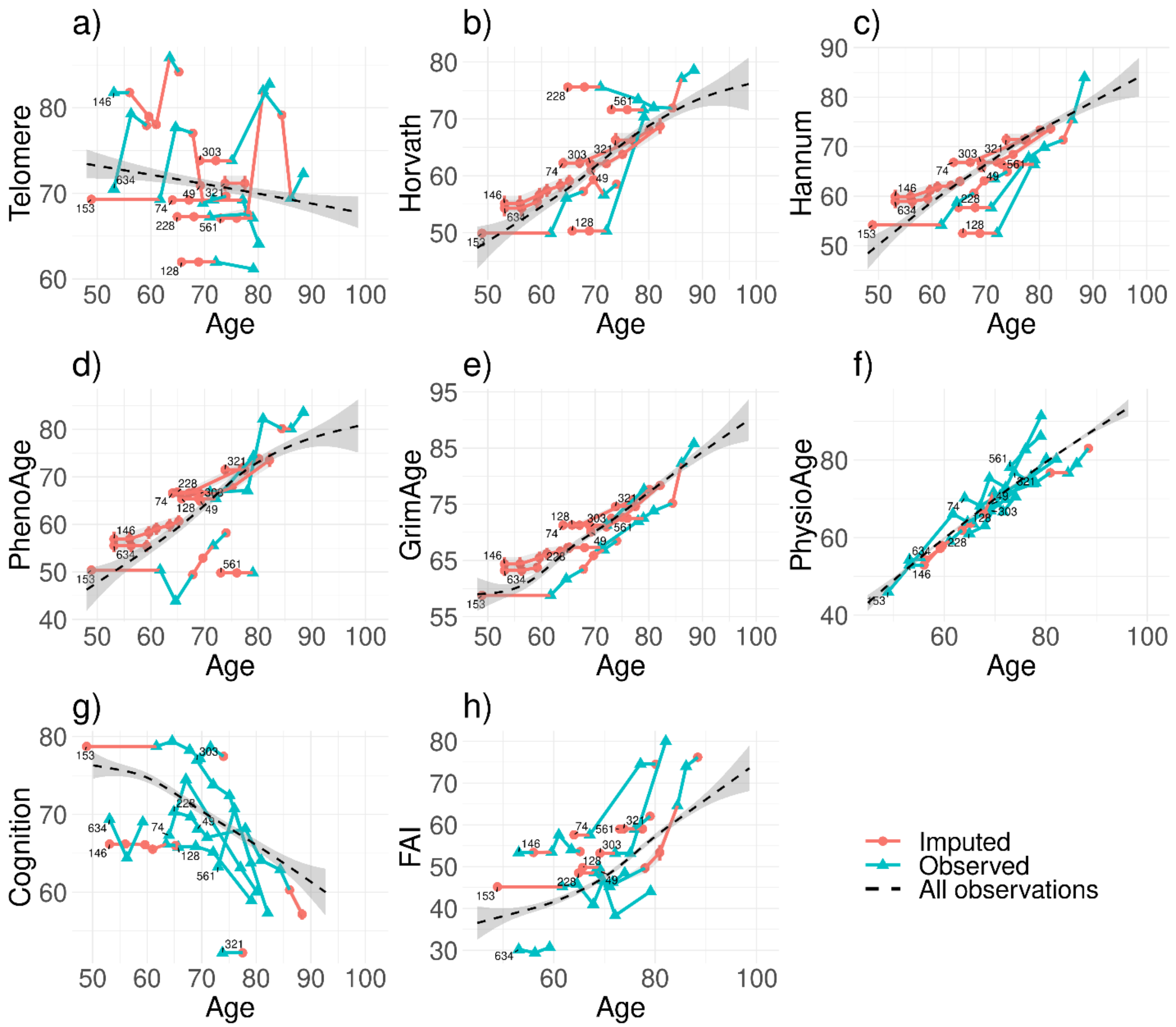

Figure S4: **Imputation quality —sample trajectories.** Sample trajectories for a random subset of 10 individuals (numbered labels). Timepoints which were measured are represented as blue triangles, imputed values are red points (with error bars). The small error bars demonstrate a low sensitivity to parameter estimates. The overall population mean trajectory is included for comparison (dashed line with error band). Interpolated points look good (e.g. PhenoAge for individual 153 at age 70). Overall trajectories look reasonable and follow the overall population trend. Overall population trend is best fits from a cubic spline additive model using the MGCV package with default degrees of freedom[58]. Error bars of imputed values are bootstrap standard deviation.

We investigate the possibility that the imputation strategy could lead to an over or under–estimate of the interaction strengths. In Figure S5 we compare the missingness to the estimated network. We see no evidence of imputation-induced bias. In Sensitivity Analysis we observed that GrimAge and PhysioAge connections were consistent across age strata, which further supports insensitivity to imputation, since there is a strong age-dependence to the missingness (Figure S19 and S20).

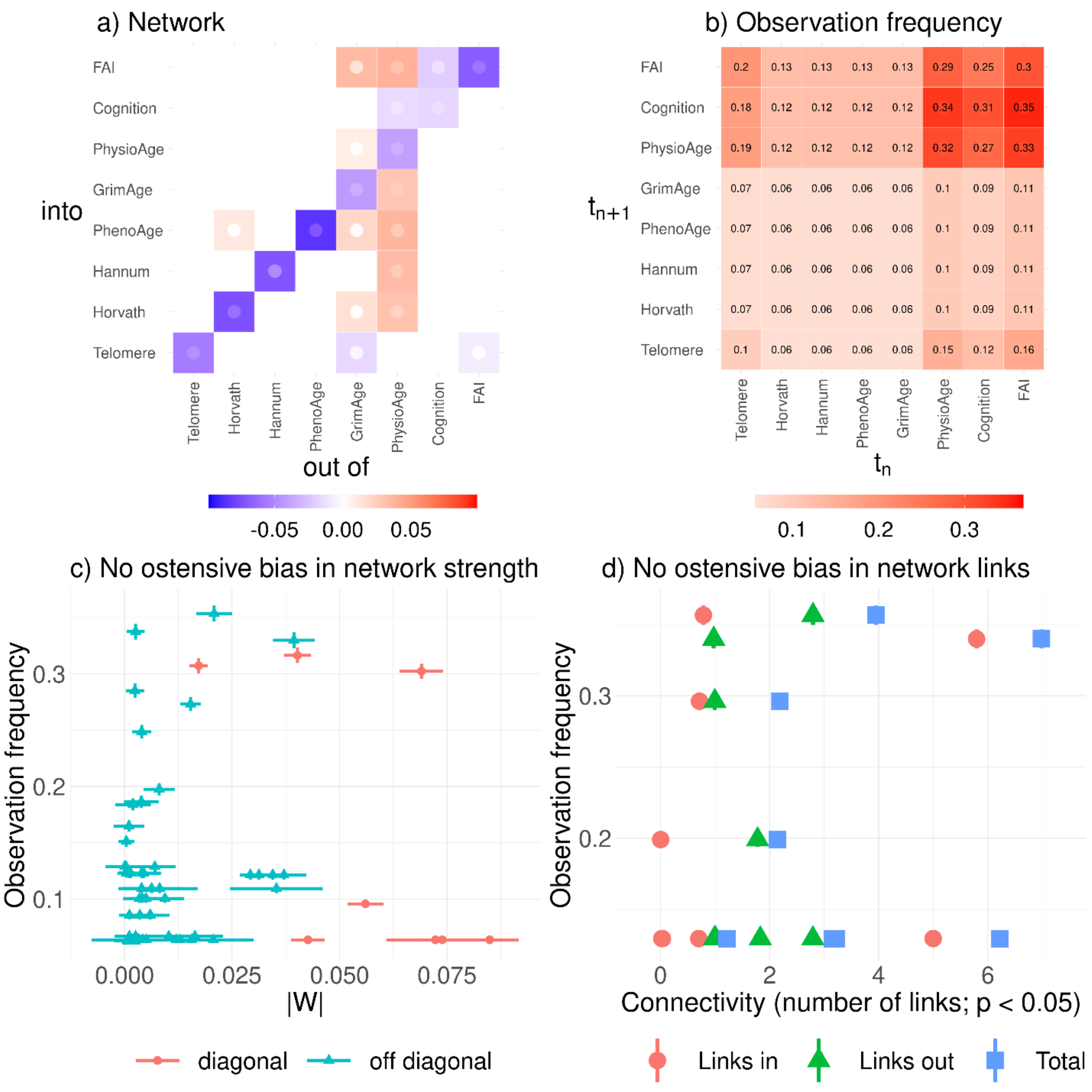

Figure S5: **Does the missingness bias the network estimation? a)** network estimate reproduced from Figure 1. Inner point is limit of 95% CI closest to 0: point is most visible for the least significant tiles. **b)** mutual observation frequency: each tile indicates how often x and y variables were both observed such that they could contribute to the log-likelihood. **c)** observation frequency versus magnitude $|\boldsymbol{W}|$ shows no correlation in diagonal nor off diagonals suggesting

that level of missingness does not affect the fit. **d)** significant links from **c** confirm that there is no ostensive bias in the network due to missing values. Missing values were imputed as described in text.

We consider alternative missing data handling methods. We can use available case analysis using our previously derived iterative estimators[23]. These estimators are based on pairwise covariances and therefore we were able to estimate using all available pairs of values. The subsequent estimated network is reported in Figure S7. The network has a similar structure to the imputed case, in particular the same dominant central nodes: GrimAge and PhysioAge.

We also consider a different imputation algorithm. The use of multiple imputation permits us to estimate the uncertainty in imputed values using Rubin's rules[59]. By (correctly) estimating the error in the imputed values we can do no harm since any under-estimate in effect sizes introduced by the imputation should be compensated for by this increased error. That is, multiple imputation is "proper" so long as it is unbiased and has realistic error estimates[60]. We consider the effects of both missed measurements and dropout. The model we employ is a multilevel linear model where individuals are allowed their own slope, permitting individualized (linear) trajectories[60] (2l.pan[61]). When we imputed dropout we assumed those individuals' age trajectories continued forward until the end of the study, such that each individual had 9 timepoints. We used MICE (multiple imputation by chained equations) version 3.13.0 for R[61]. We included age and the FI in the imputation. We imputed each individual 20 times, since this number should be large enough to capture the underlying distribution of values while still being manageable. Each of the 20 imputations produced a separate dataset which we fit using the methodology outlined in the main text. Where specified we imputed dropout individuals, otherwise we exclude all values past dropout date. The parameters from each of these datasets, including the network estimates, were then pooled using Rubin's rules. This provides us with both an average estimate of the 'true' parameter values as well as an uncertainty estimate which includes stochastic effects via the bootstrap and uncertainty in the imputation via the multiple imputations.

We investigated the imputation quality of MICE both as individual trajectories (not shown) and at the population-level, Figure S6 (includes dropout imputation). In both cases the imputed values looked very similar to the observed values with similar mean and dispersion. The multiple imputations gave realistic error estimates for the range of possible values which a missing value could take. While this supports the conclusion that the MICE imputation better estimates the uncertainty in the parameter estimates, in does not necessarily indicate a better point estimate for the central (mean) parameter values. We know that the missing values are more likely to occur in the older individuals which suggests that the true values of the missing values should be older than the observed values. We don't see this effect in the MICE imputation (Figure S6), whereas we do with the expectation-maximization imputation (Figure S3): this can be seen by following the trendlines which indicate a preference to impute data as older in Figure S3 whereas the trendlines coincide in MICE. Overall, however, the MICE imputation looked good.

We present the network estimates for four missing data handling methods in Figure S7 and the associated significance scores in Figure S8. We expect that the available case will under-estimate any effects and may be biased due to the non-random missingness (i.e. older, frailer individuals were more likely to be missing data). We further expect that the main result, which used expectation-maximization, will likely under-estimate errors and hence may over-estimate effect sizes. The MICE imputed networks should compensate for errors in the imputation process and hence should give realistic estimates for the true networks. Across the approaches we observe that GrimAge and PhysioAge are consistently the central nodes with the largest outgoing connections. This supports our primary interpretation of the network. We do see apparent differences between the diagonal strength in the main result and MICE but this is in the epigenetic ages and hence does not affect any of our conclusions. The available case notably showed a weaker role for PhysioAge and stronger role for Cognition in contrast to the other imputation strategies — which had a consensus central role for PhysioAge and minimal role of Cognition.

We also confirm that the natural variables yielded by MICE have similar properties to those in the main text. While we confirm this, it is worth noting that MICE-imputed values yielded natural variables which differ in a few interesting ways: Figure S9 vs Figure 3. Looking at the FI row we clearly see a strong propensity for health information to compress into the first few dimensions. While this effect is also seen in Figure 3, it is much cleaner in the MICE-imputed data which includes error estimates and an associated p-value cutoff at 0.05. The splitting of the first and second eigenvalues is also much stronger in the MICE data, which leads to bigger differences in the first and second natural variables.

Finally, we offer a note on our choice of imputation strategy. While MICE has the great advantage of permitting us to estimate the uncertainty in imputed values, we are concerned that it may have a stronger bias than our main choice (expectation-maximization). We had expected imputed individuals to look older since older individuals tended to have more missing data. We did see such an effect in the main approach (Figure S3) but did not in MICE (Figure S6). We also saw in Figure S20 that the age-stratified networks estimated by MICE were much more similar across age cohorts than were the main approach (Figure S19), which could indicate that the older individuals were made to look more like young individuals by MICE. We see two shortcomings of MICE: (1) it uses Fully Conditional Specification which is not formally self-consistent[60] as opposed to our main approach, and (2) due to the missingness bias, younger individuals had more data and may have therefore had more weight in the imputation model learned by MICE. While this latter point could also apply to our main result, the fact that we use an auto-regressive model should make imputed values less sensitive to parameter values. Regardless, both imputation strategies yielded similar results and we can get some idea of the broader uncertainty in our results based on how the two compare.

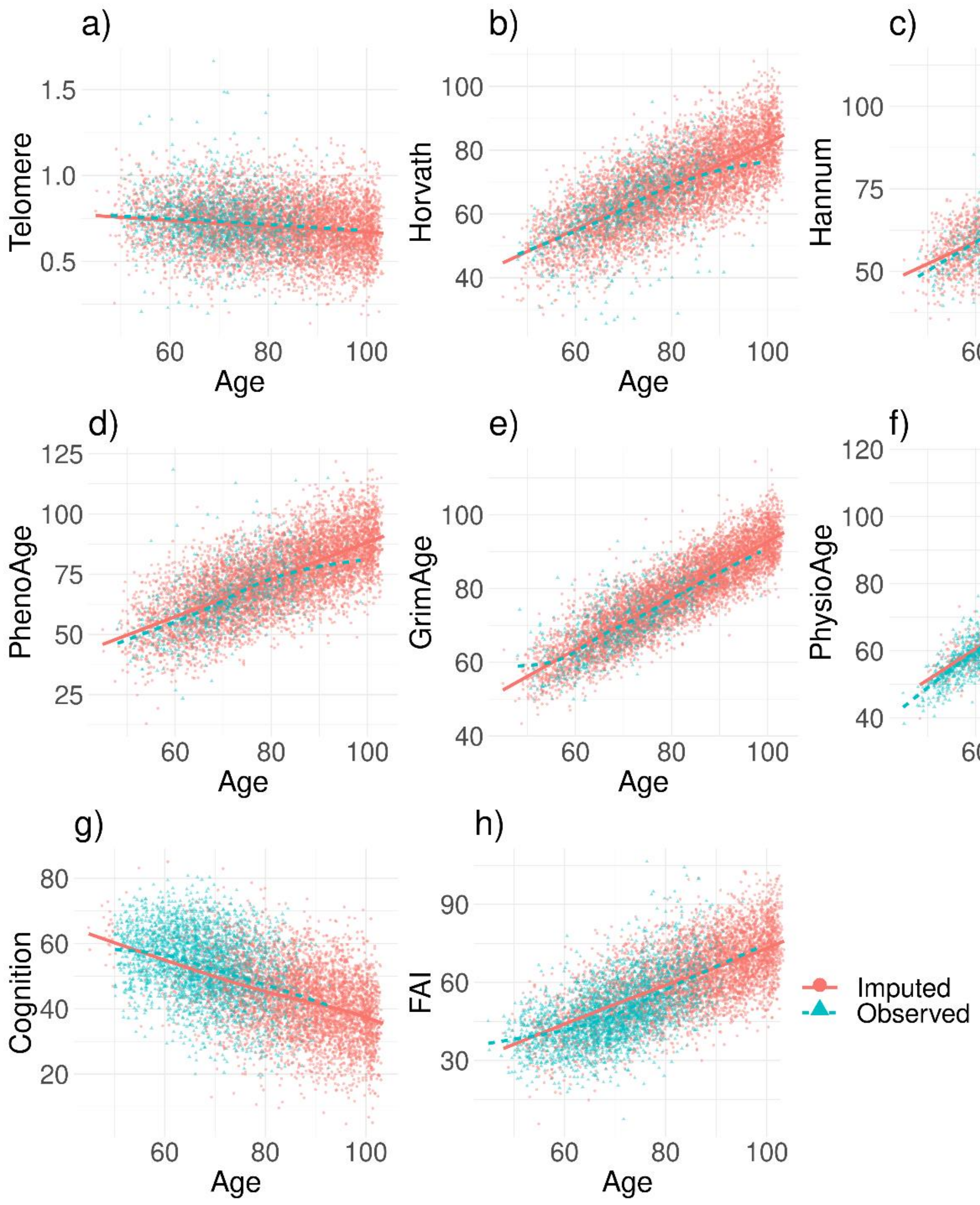

Figure S6: **MICE imputation quality (multilevel linear model with individual slope).** Population level data for all 8 BAs. Each individual was imputed 20 times and hence has 20 imputed values for each datum. As with Figure S3 we look for similar dispersion and trend with the imputed (red dots) and observed (blue triangles) values. In contrast to Figure S3, we have also imputed values after dropout which we then optionally removed (hence the values extend to older ages). We see nearly identical trends between the imputed and observed values: there is a best fit line for each, on each plot, but they typically overlap. We also see realistic dispersion. These are all good signs. Note, however, that we know that the missing values are biased towards older individuals so we would expect that the best fit line for the imputed should be a bit higher than the observed – which we see in Figure S3 but not here. It is therefore unclear which of the two offers the better imputation. The advantage of MICE is that it quantifies the uncertainty in the imputation using Rubin's rules.

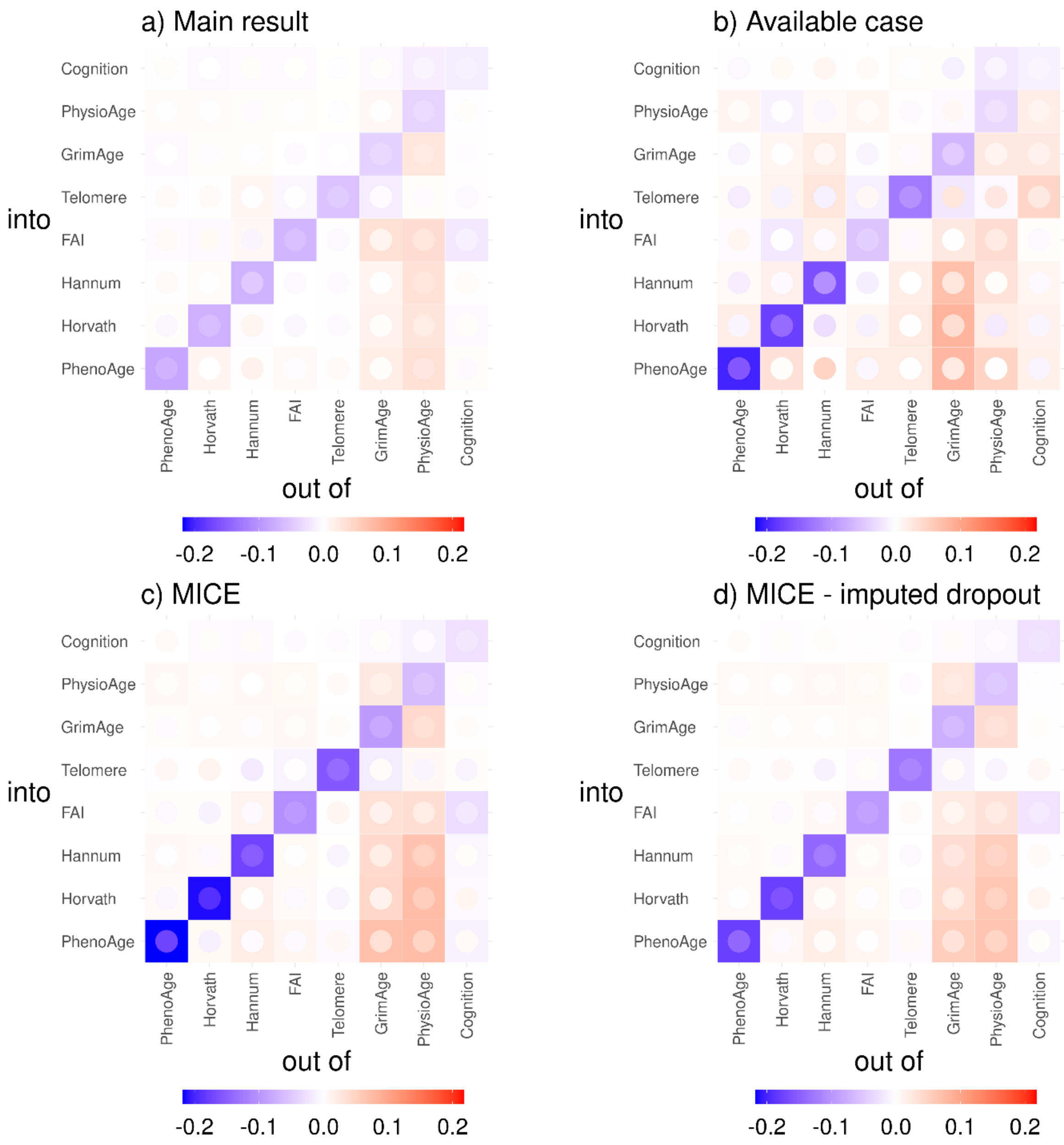

Figure S7: **Effect of missing data handling on network estimate.** We compare 4 missing data handling strategies (**a-d**). In all cases we see strong outgoing links from GrimAge and PhysioAge, supporting our main result which implicates them as the central nodes. We expect that the MICE imputation **(c)** will give the best confidence estimates since it utilizes multiple imputation and Rubin's rules to estimate the uncertainty in imputation. By including dropped

individuals, **d)**, we can also surmise the effect that censorship has on the network — which is clearly negligible (**c** and **d** look almost identical). Inner point is limit of 95% CI closest to 0: point is most visible for the least significant tiles; if point is opposite colour to tile then element is not significant. See Figure S8 for link significance (z-scores).

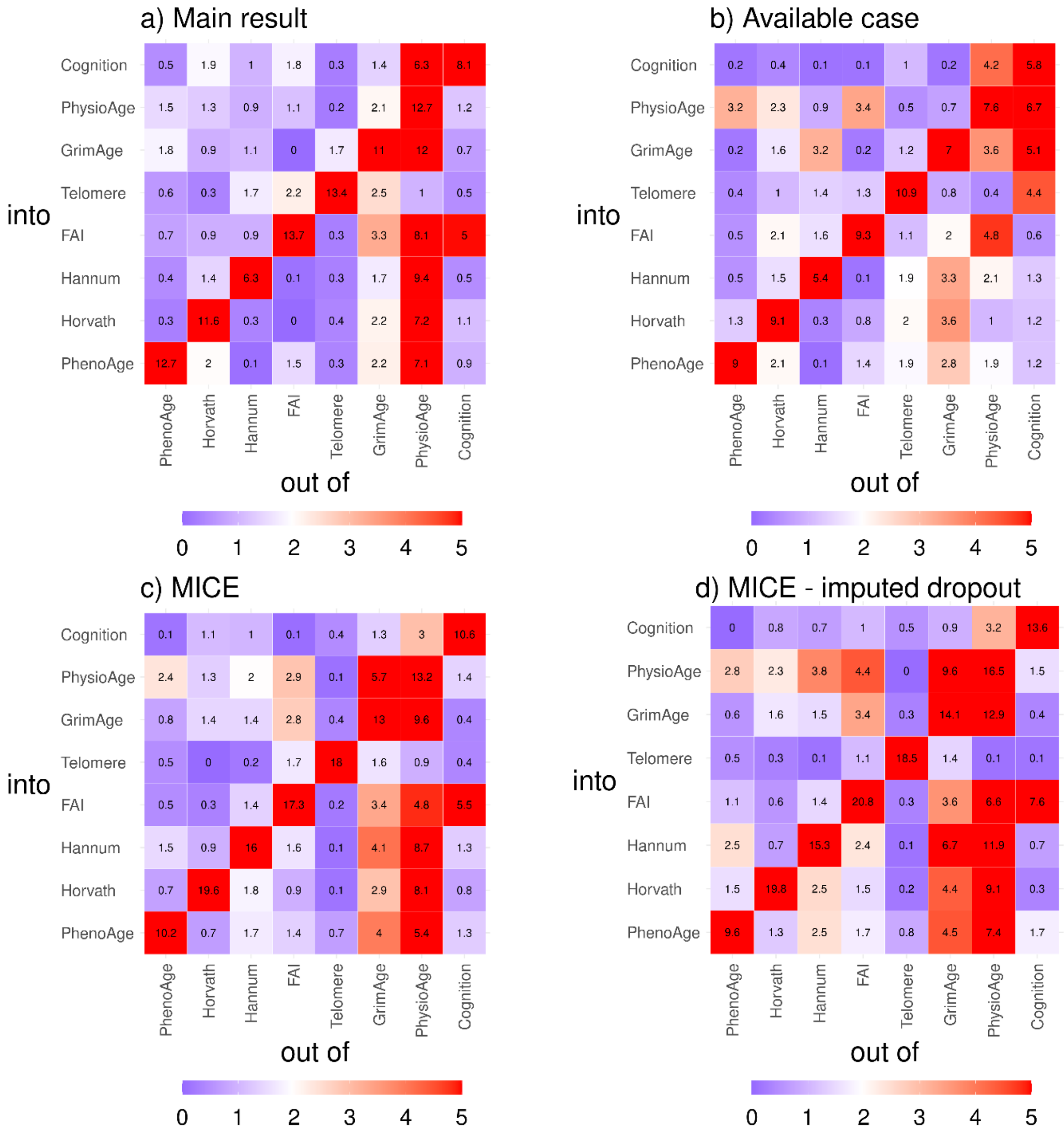

Figure S8: **significant network links by missing data handling method (z-scores)**. Blue are non-significant, red and white are significant at $p < 0.05$ ($z > 1.96$). Standard errors were estimated by bootstrap (100 repeats). Observe the very high significance of the links outgoing from PhysioAge, supporting its central role in the network.

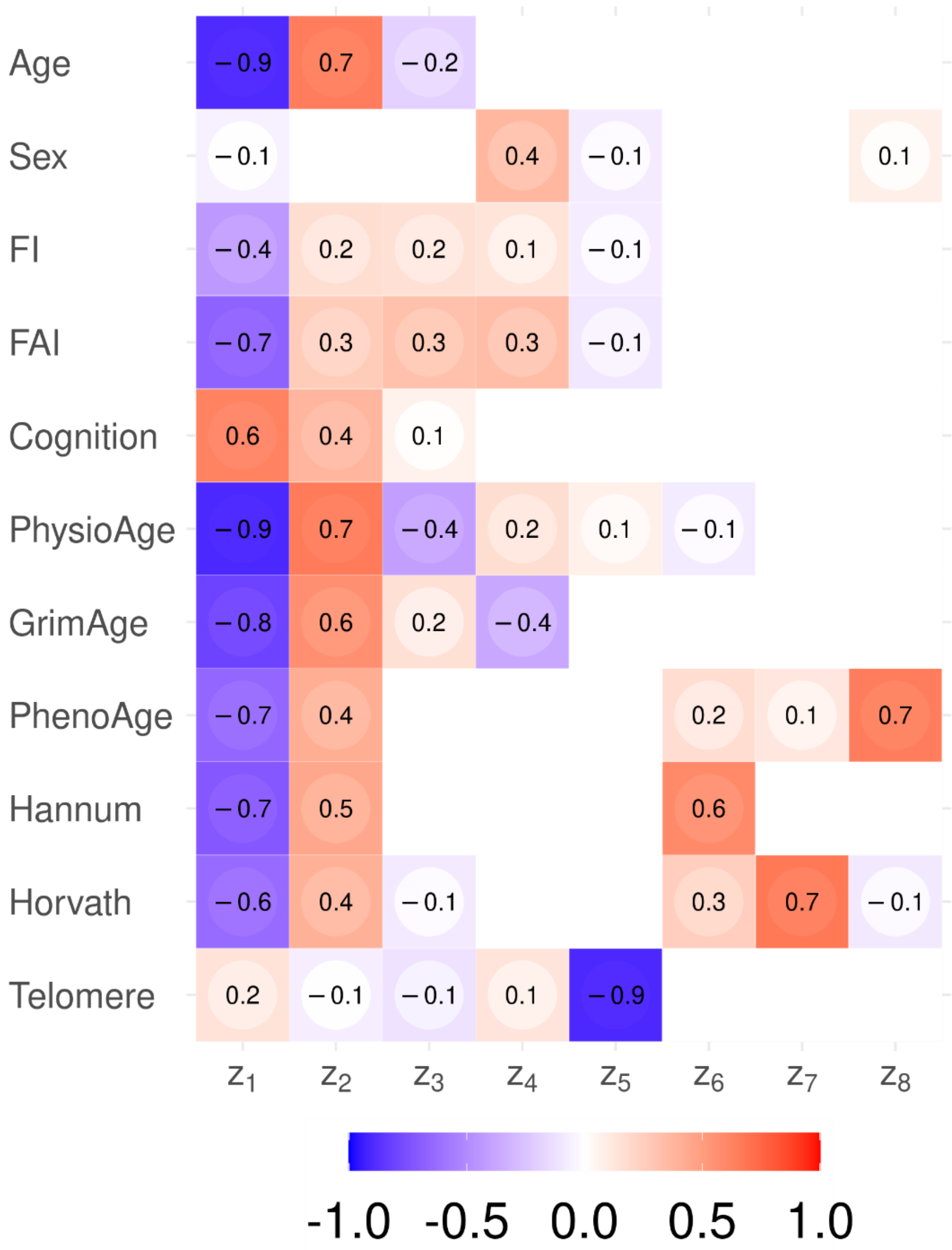

Figure S9: **Natural variable correlates — MICE**. Spearman correlations between the natural estimates and the original variables for the MICE-imputed data. Errors were estimated by 100 repeat bootstrap. Non-significant tiles by z-scores are whited-out ($p > 0.05$). Inner point is limit of 95% CI closest to 0: point is most visible for the least significant tiles.

**Estimation**

We previously reported an iterative estimator as well as a diagonal-space weighted linear regression estimator for the Stochastic Finite-difference (SF) model[23]. Here we show that the linear regression estimator applies more generally, including for all invertible $\boldsymbol{W}$. The linear estimator is always valid for $\boldsymbol{W}$ but can only estimate $\boldsymbol{\Lambda}$ and $\vec{\mu}_0$ if $\boldsymbol{W}$ is invertible ($\boldsymbol{W}$ can be stable or unstable but cannot have any eigenvalues equal to 0). Usually this is not a problem since real data is unlikely to have an eigenvalue of exactly 0, but we can nevertheless circumvent the problem entirely by pre-processing using principal component analysis (PCA), since this will allow us to drop reduced-rank terms (i.e. eigenvalues close to 0). The final parameters can then be mapped back into observed space using the PCA transformation.

Recall that our SF model is written as

$$\begin{aligned} &\vec{b}_{n+1} = \vec{b}_n + \boldsymbol{W}\Delta t_{n+1}\left(\vec{b}_n - \vec{\mu}_n\right) + \vec{\epsilon}_{n+1} \\ &\vec{\epsilon}_{n+1} \sim N(0, \boldsymbol{\Sigma}|\Delta t_{n+1}|) \\ &\vec{\mu}_n \equiv \vec{\mu}_0 + \boldsymbol{\Lambda}\vec{x}_n \end{aligned} \tag{S8}$$

which can be rewritten

$$b_{jn+1} - b_{jn} = \sum_k W_{jk}\,\Delta t_{n+1} b_{jn} - \sum_{k,l} W_{jk}\,\Delta t_{n+1}\Lambda'_{jl} x_{ln} + \epsilon_{jn+1} \tag{S9}$$

for each BA, where $\boldsymbol{\Lambda}'$ includes $\vec{\mu}_0$ by inventing a new $x_n = 1$. This can be rewritten in a more revealing way as

$$\Delta b_{jn+1} = \boldsymbol{W}_{j\cdot}\vec{\alpha}_n + \mathbf{A}_{j\cdot}\vec{\beta}_n + \epsilon_{jn+1} \tag{S10}$$

where $\alpha_{jn} \equiv \Delta t_{n+1} b_{jn}$, $\beta_{jn} \equiv \Delta t_{n+1} x_{jn}$, $\mathbf{A} \equiv -\mathbf{W}\boldsymbol{\Lambda}$ and $\Delta b_{jn+1} \equiv b_{jn+1} - b_{jn}$ ($\boldsymbol{W}_{j\cdot}$ is the $j$th row of $\boldsymbol{W}$). Eq. (S10) is a (weighted) linear regression equation (note: because of the noise you must weight each entry by $1/|\Delta t_{in+1}|$ as well[23]). These regression equations estimate $\boldsymbol{W}$ and the product $\mathbf{A} = -\boldsymbol{W}\boldsymbol{\Lambda}$, hence we must be able to invert $\boldsymbol{W}$ to estimate $\boldsymbol{\Lambda}$ using linear regression ($\det(\boldsymbol{W}) \neq 0$). We used Eq. (S9) to estimate our model parameters $\boldsymbol{W}$ and $\boldsymbol{\Lambda}$ ($\vec{\mu}_0$ was estimated via $\boldsymbol{\Lambda}$ by inventing a constant $x_n \equiv 1$). The noise, $\boldsymbol{Q} = \boldsymbol{\Sigma}^{-1}$ was estimated using the model residual[23].

Note that although the error terms are coupled across variables, this does not affect estimation.

**Proof**

Consider the (equivalent) simplified form of our problem,

$$\vec{y} = \boldsymbol{W}\vec{v} + \vec{\epsilon} \tag{S11}$$

where $\vec{\epsilon} \sim N(0, \boldsymbol{\Sigma})$. Because $\boldsymbol{\Sigma}$ is a covariance matrix it must be positive definite and hence there exists a transformation $\boldsymbol{P}$ which diagonalizes $\boldsymbol{P}^{\boldsymbol{T}}\boldsymbol{\Sigma}\mathbf{P} = \boldsymbol{\Gamma}$ with $\mathbf{P}^{\mathbf{T}} = \mathbf{P}^{-\mathbf{1}}$ for diagonal $\boldsymbol{\Gamma}$. Hence $\boldsymbol{P}^{\boldsymbol{T}}\vec{\epsilon} \sim N(0, \boldsymbol{\Gamma})$. This means we can transform Eq. (S11) such that the equations decouple and can safely be estimated independently,

$$z_j = \widetilde{\boldsymbol{W}}_{j\cdot}\vec{u} + \gamma_j \tag{S12}$$

Where $\widetilde{\boldsymbol{W}} \equiv \boldsymbol{P}^{\boldsymbol{T}}\boldsymbol{W}\boldsymbol{P}$, $\vec{\gamma} \equiv \boldsymbol{P}^{\boldsymbol{T}}\vec{\epsilon}$, $\vec{z} \equiv \boldsymbol{P}^{\boldsymbol{T}}\vec{y}$ and $\vec{u} \equiv \boldsymbol{P}^{\boldsymbol{T}}\vec{v}$. The least squares estimator is[62]

$$\widetilde{\boldsymbol{W}}_{j\cdot} = \left(\sum_i \vec{u}_i\vec{u}_i^T\right)^{-1} \sum_i \vec{u}_i z_{ij}$$

$$\widetilde{\boldsymbol{W}} = \left(\sum_i \boldsymbol{P}^{\boldsymbol{T}}\vec{v}_i{\vec{v}_i}^T\boldsymbol{P}\right)^{-1} \sum_i \boldsymbol{P}^{\boldsymbol{T}}\vec{v}_i\vec{y}_i^{\boldsymbol{T}}\boldsymbol{P} \tag{S13}$$

$$\widetilde{\boldsymbol{W}} = \boldsymbol{P}^{\boldsymbol{T}}\left(\sum_i \vec{v}_i{\vec{v}_i}^T\right)^{-1} \sum_i \vec{v}_i\vec{y}_i^T\,\boldsymbol{P} \tag{S14}$$

$$\boldsymbol{W} = \left(\sum_i \vec{v}_i{\vec{v}_i}^T\right)^{-1} \sum_i \vec{v}_i\vec{y}_i^T, \tag{S15}$$

which is exactly the ordinary least squares estimator of Eq. (S11).

**QED**

## Additional Results

### Model Selection

Here we provide additional results that were not included in the main text for want of space. First we demonstrate that our model works. In Figure S10 we compare 3 variants of our model to simply carrying forward the previous value to predict the future value (equivalent to $\boldsymbol{W} = \boldsymbol{0}$). The fully flexible model is FullW, whereas the other models are simplified variants to reduce overfitting, specifically: assume that $\boldsymbol{W}$ is diagonal (DiagW) or assume that $\boldsymbol{W}$ is diagonal in PC-space (SymW, where PCA forces $\boldsymbol{W}$ to be symmetric because PCA is an orthogonal transformation). In both FullW and SymW we first performed a PCA transformation to avoid collinearity issues and potential fit problems due to reduced-rank $\boldsymbol{W}$ (the transformation was learned using timepoint 1). We use 632-corrected root mean squared error (RMSE) and mean absolute error (MAE) to quantify error (lower is better); e.g.[62] $\text{RMSE}_{632} \equiv 0.632 \cdot \text{RMSE}_{\text{test}} + 0.368 \cdot \text{RMSE}_{\text{train}}$ (632-correction reduces bias[23]). We observe that all of our models out-perform forward carry (in which all values are carried forward unchanged to the next timepoint). Additionally, FullW and SymW out-performed DiagW, indicating that an interaction network is supported by the data.

Whereas the FullW model includes directed links, SymW is a parsimonious model that permits only undirected links. While FullW appeared to have a lower MAE, the RMSE was the same as SymW — within error. To break the tie, we consider a simplified accuracy measure: which model best predicts worsening. In Figure S11 we present the AUC[33] for correctly predicting worsening. We observe that a causally unambiguous model (FullW) out-performs the symmetrical $\mathbf{W}$ (SymW) at $p = 0.1$ (Delong test[33]). Hence, for both the MAE and the AUC we find FullW out-performs SymW at 68% confidence (non-overlapping error bars). If the two metrics (MAE and AUC) are independent we have $p = 0.1 \cdot (1 - 0.68) = 0.04$ which is significant at 95% confidence. Given the evidence, we selected FullW as the best model.

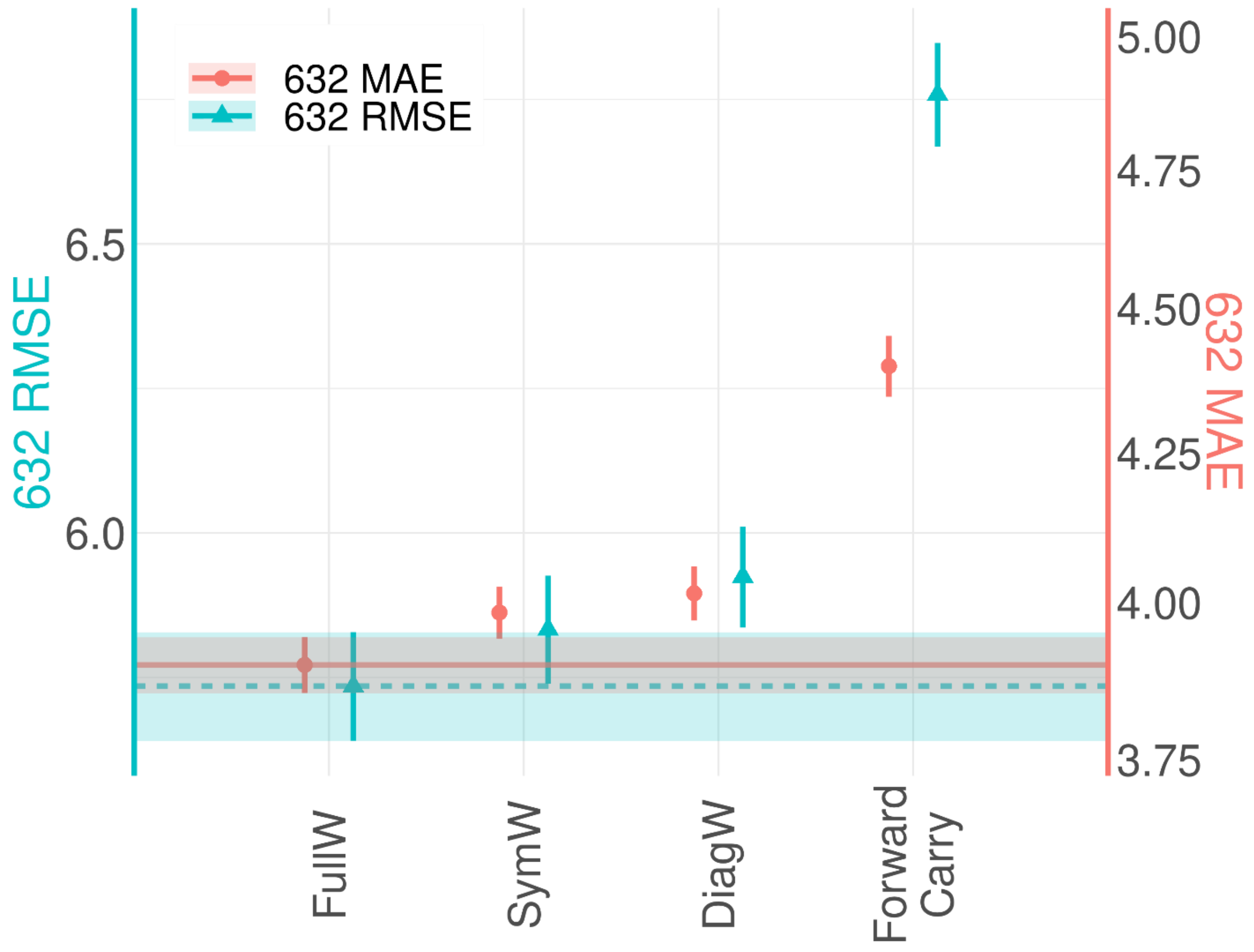

Figure S10: **Model error, lower is better.** Our model easily out-performs forward carry. The diagonal $\boldsymbol{W}$ (DiagW) performs worse than the full $\boldsymbol{W}$ (FullW) for both MAE and RMSE (bands). The symmetric $\boldsymbol{W}$ (SymW), on the other hand, performs worse in MAE but not RMSE. Scale is years of age. Error bars are standard error.

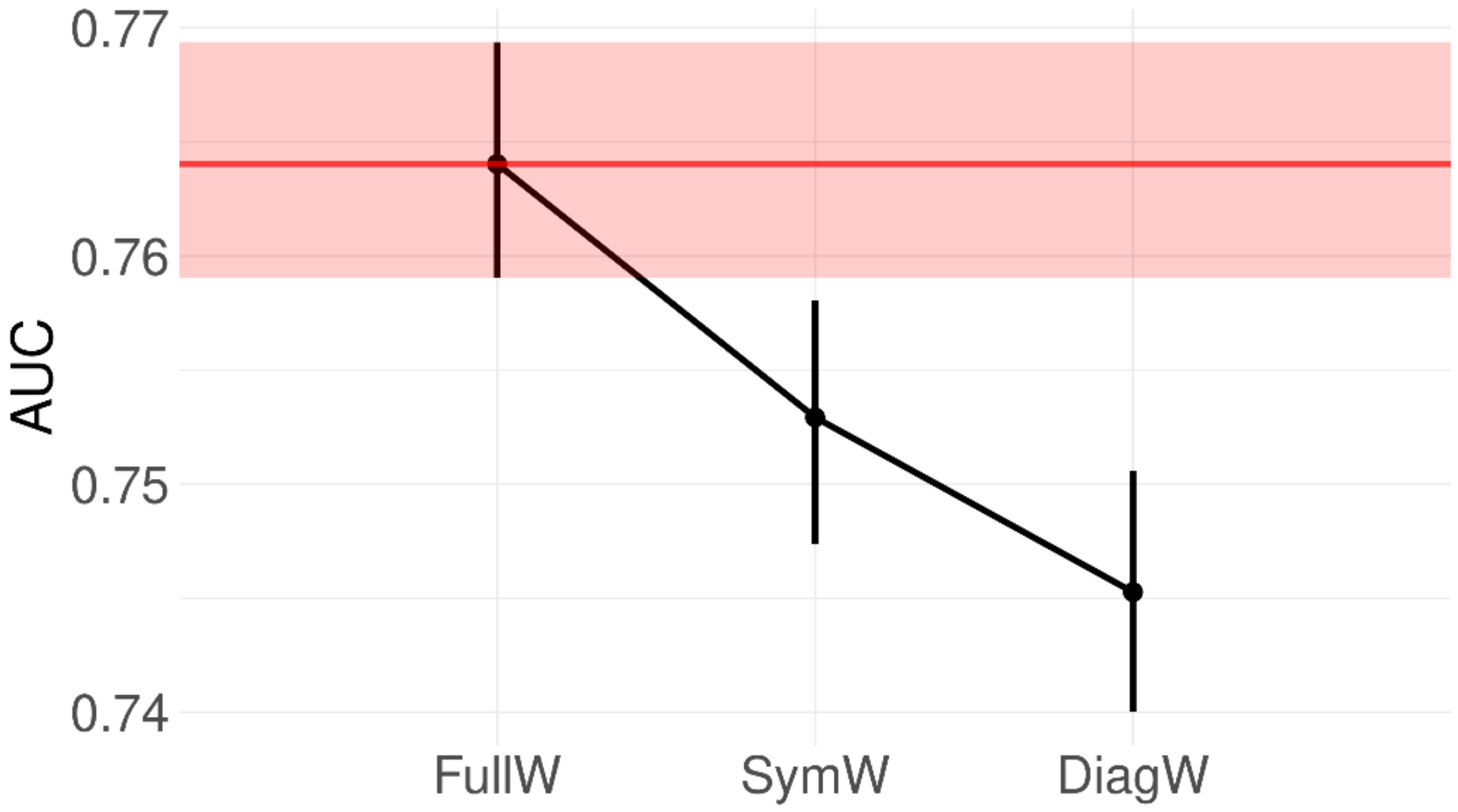

Figure S11: **Worsening detection model performance**, higher is better (0.5: guess, 1: perfect). The symmetric $\boldsymbol{W}$ model (SymW) performed worse than the model which includes a fully flexible $\boldsymbol{W}$ (FullW), $p = 0.1$ (Delong test[33]). The diagonal model (DiagW) performed worse than FullW, $p = 0.01$ (Delong test[33]). Forward carry did not perform better than a guess (AUC: $0.501 \pm 0.006$, not shown). Error bars are standard errors using N=2000 resample bootstrapping[33]. Individuals worsened in 60.0% of measurements (they improved in 40.0% of measurements).

**Model Parameters**

The full model estimates an unconstrained matrix of network weights, $\mathbf{W}$, together with equilibrium state $\vec{\mu}_n$ which depends on sex, and a positive-definite stochastic noise term, $\mathbf{\Sigma}$. Whereas $\mathbf{W}$ is presented in the main text, $\vec{\mu}_n$ (Table S1) and $\mathbf{\Sigma} = \mathbf{Q}^{-1}$ (Figure S12) are presented below. The exact model parameters are provided in CSV file format in the GitHub repository, these can be used to simulate interventions https://github.com/GlenPr/stochastic_finite-difference_model.

The equilibrium state is determined by $\vec{\mu}_n$, although the system converges if and only if $\lambda_k < 0$ (and at a rate of $-\lambda_k$). In Table S1 we present the coefficients of $\vec{\mu}_n$, including the intercept, $\vec{\mu}_0$ and the sex-dependence ($\Lambda_{sex}$). The sex-dependence is weak, reflecting that the model isn't sensitive to starting positions for each individual at their baseline (the sex dependence appears to be small and due to starting position[9]). Observe that *in all cases* the $\mu_{j0}$ is much different from the population age range (45-88) and always in the risk direction. This ensures that all individuals drift towards worsening BAs as they move towards the equilibrium position. The equilibrium positions are far enough away that they will never be reached since the individuals will die long before that happens. The errors are large, ostensibly because it is important that $\mu_{j0}$ is far away but less important exactly how far away. Figure S14 illustrates.

Table S1: equilibrium positions, $\mu_0$ and $\boldsymbol{\Lambda}$ (units of years).

| Biological age | Risk Direction[1] | $\mu_0$ | $\Lambda_{sex}^{(2)}$ |
|---|---|---|---|
| Telomere | Down | $35 \pm 24$ | $4 \pm 3$ |
| Horvath | Up | $127 \pm 38$ | $-5 \pm 4$ |
| Hannum | Up | $130 \pm 35$ | $-5 \pm 4$ |
| PhenoAge | Up | $134 \pm 41$ | $-4 \pm 4$ |
| GrimAge | Up | $136 \pm 37$ | $-6 \pm 4$ |
| PhysioAge | Up | $149 \pm 43$ | $-4 \pm 5$ |
| Cognition | Down | $-122 \pm 107$ | $10 \pm 12$ |
| FAI | Up | $190 \pm 82$ | $-4 \pm 9$ |

(1) Direction of change with increasing chronological age.

(2) Male: 0, female: 1 (coefficient modifies only females).

Our model captures stochastic effects via the noise term, $\boldsymbol{\Sigma}$. In Figure S12 we present the normalized $\boldsymbol{\Sigma}$, which reflects correlations between the biological ages in their response to stochasticity: stressors, individual variabililty, and non-linearities. We observe three self-evident blocks, representing three scales: telomere noise does not correlate with the other biomarkers, epigenetic ages all mutually correlate strongly along with PhysioAge, and finally Cognition and FAI correlate mutually. These blocks suggest that stochastic effects are not shared across scales. The exception appears to be PhysioAge and GrimAge which both couple to Cognition and FAI. Between $\boldsymbol{W}$, which is related to the resilience via recovery rate, and $\boldsymbol{\Sigma}$, which is related to robustness via stressor effects, we consistently observed central roles for PhysioAge and GrimAge: for connecting, and potentially driving, the changes observed in the other BAs.

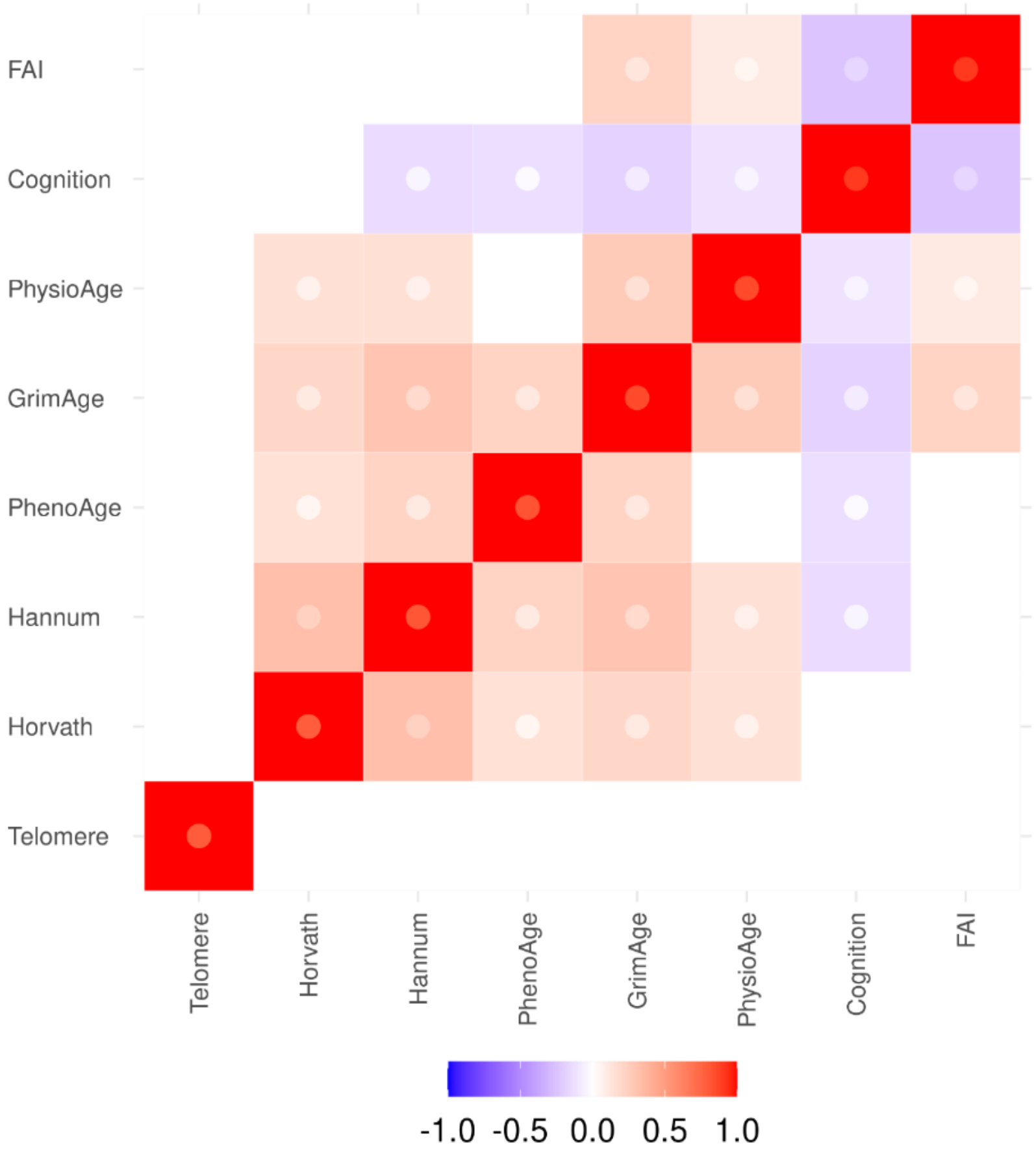

Figure S12: **Normalized noise matrix, $\boldsymbol{\Sigma}$**. Normalized by diagonal strength. Note the overlapping, block-diagonal structure. The blocks suggest common domains while the overlapping ages may represent feedbacks between domains, i.e. GrimAge, PhysioAge and, to an extent, Cognition. Inner point is limit of 95% CI closest to 0: point is most visible for the least significant tiles. Non-significant tiles are whited-out ($p > 0.05$).

### Natural Variables

Eigen-decomposition of the interaction network, $\boldsymbol{W}$, yields a transformation matrix of eigenvectors ($\boldsymbol{P}$ in Eq. (S7)). The eigenvector represents canonical coordinates which decouple the mean-interactions, greatly simplifying the dynamics (they now satisfy Eq. (3)). The eigenvector transformation matrix, $\boldsymbol{P}$, can be used to generate scores for each of the $i$ individuals using their data vector, $\vec{b}_{in}$, similar to the way in which PCA generates PC scores. This generates natural aging variables, $\vec{z}_{in} \equiv \boldsymbol{P}^{-1}\vec{b}_{in}$, which are aggregated BAs. The age-dependence of the natural variables are plotted in Figure S13; for comparison, the BAs are plotted in Figure S14.

It is noteworthy that the strongest correlations with age are present in the lowest $z_j$, suggesting that the aging phenomenon is concentrated into the slowest recovery dimensions (recall $z_j$ has the $j$th slowest recovery). Also observe the positions of the equilibrium positions, $\mu_j$. In the BA-picture (Figure S14), $\mu_j$ was always far away in the risk direction, ensuring that each BA drifted continuously for each individual during their lifespan. In the z-picture, equilibrium is quickly reached for all of the fast recovering $z_j$, with the slower $z_3$ equilibrating around age 80 and $z_1$ and $z_2$ never reaching equilibrium. We previously observed this same phenomenon for health biomarkers[23]. This has two effects: (1) the mean continues to drift indefinitely, causing the slow $z_1$ and $z_2$ to become increasingly dominant in the mean as individuals age, and (2) the variance will also typically continue to increase. Together this means that at advanced ages $z_1$ and $z_2$ can dominant both the mean and variance, meaning that they dominate what we observe in the BAs since they are connected via the transformation $\boldsymbol{P}\vec{z}_{in} \equiv \vec{b}_{in}$. Hence the $z_j$ are natural *aging* variables since they become increasingly simple, i.e. low-dimensional, as individuals age.

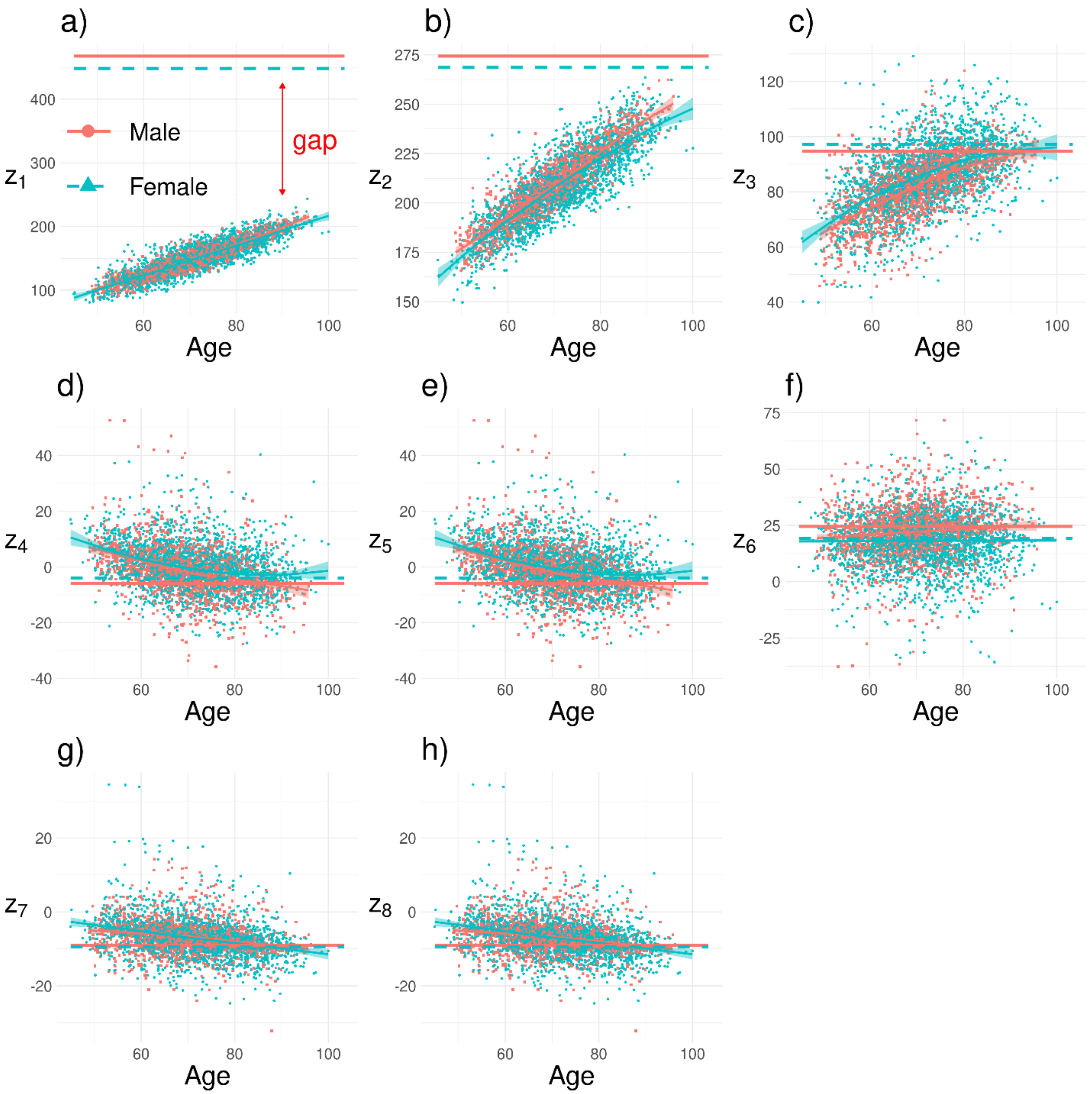

Figure S13: **natural variable age-dependencies, together with equilibrium position ($\mu_j$).** In contrast to the BAs (Figure S14), most of the natural variables equilibrate at around age 80 where they cross the horizontal equilibrium line, $\mu_j$. The exceptions are $z_1$ and $z_2$ which both have gaps between the observed values and $\mu$ (e.g. as indicated for $z_1$), indicating that they never equilibrate, instead drifting up for the entire human lifespan. This constant drift explains why $z_1$ and $z_2$ have the strongest age-dependence. We see little sex effects, primarily concentrated into $z_3$. Real components only. Note: $z_4/z_5$ and $z_7/z_8$ are conjugate pairs which differ only in their imaginary component. Lines are cubic splines from the MGCV package with default parameters[58].

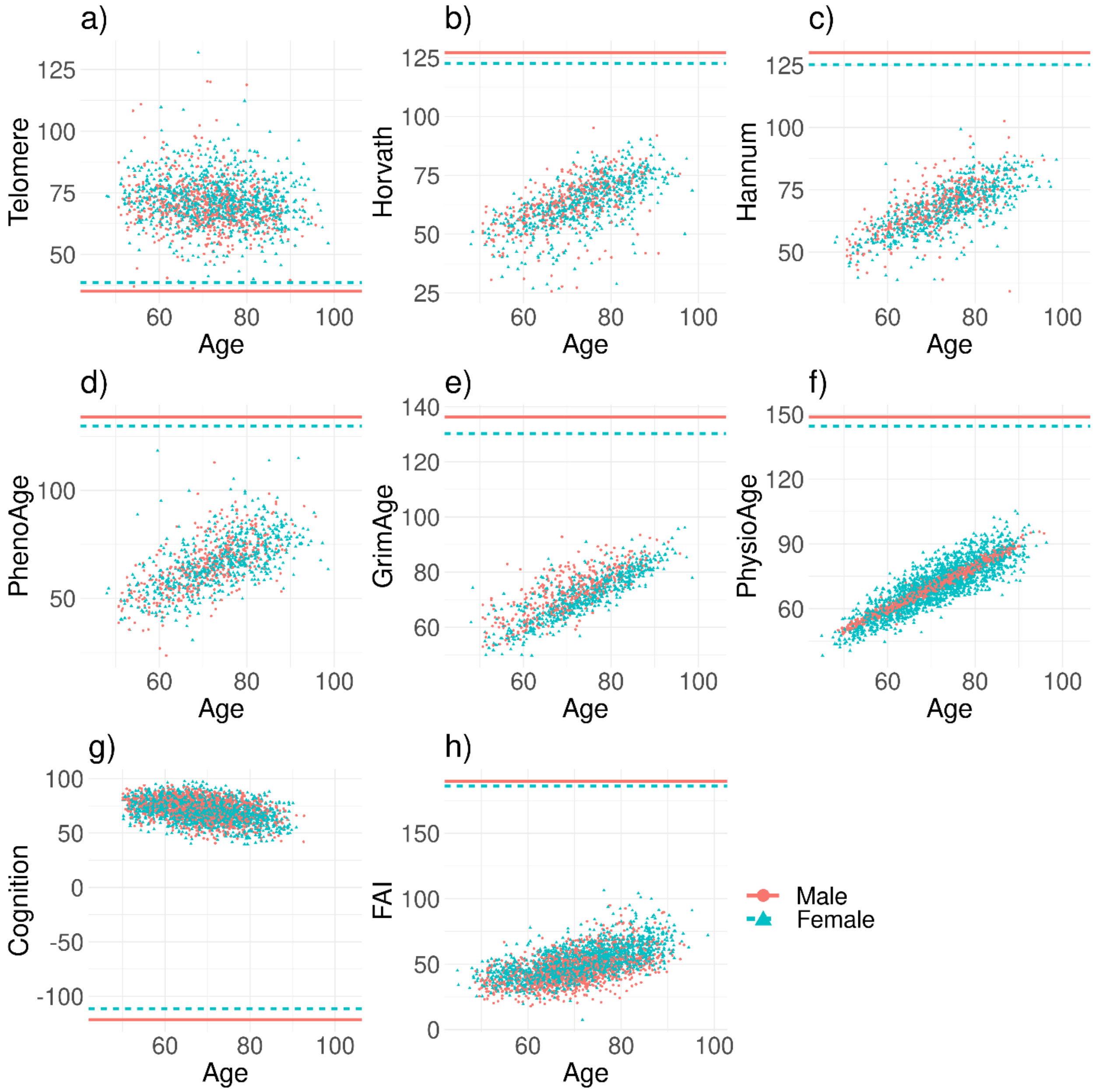

Figure S14: **biological age (BA) age-dependencies, together with equilibrium position ($\mu_j$).** Each BA is far from equilibrium (horizontal lines), and will never reach it during a normal human lifespan. Compare to the natural variables, which compress this non-equilibrium behaviour into the first two variables (Figure S13). Sex effects are relatively small (red points/solid lines vs blue triangles/dashed lines).

**Simulated Interventions**

Once we have estimated the model parameters we can simulate new data. We introduce interventions into the simulated data as instantaneous rejuvenations, emulating the rapid switching which occurs in some anti-aging interventions, e.g. dietary restriction of flies[38]. (Interventions may not look instantaneous in the figures because of the finite step size (1 year) in the simulation.) In the main text we focused on a simple intervention which rejuvenates PhysioAge by 10 years, applied at age 70. Here we considered modifications of the intervention.

First we considered instead worsening PhysioAge by 10 years. This could represent the effects of disease, for example, severe COVID is associated with persistent cognitive decline of strength comparable to aging 10 years[63]. The simulated effect is identical to that of rejuvenation, with the sign flipped, Figure S15b. We have also included a visualization of the effect of the intervention on the health trajectory of the population in Figure S15a.

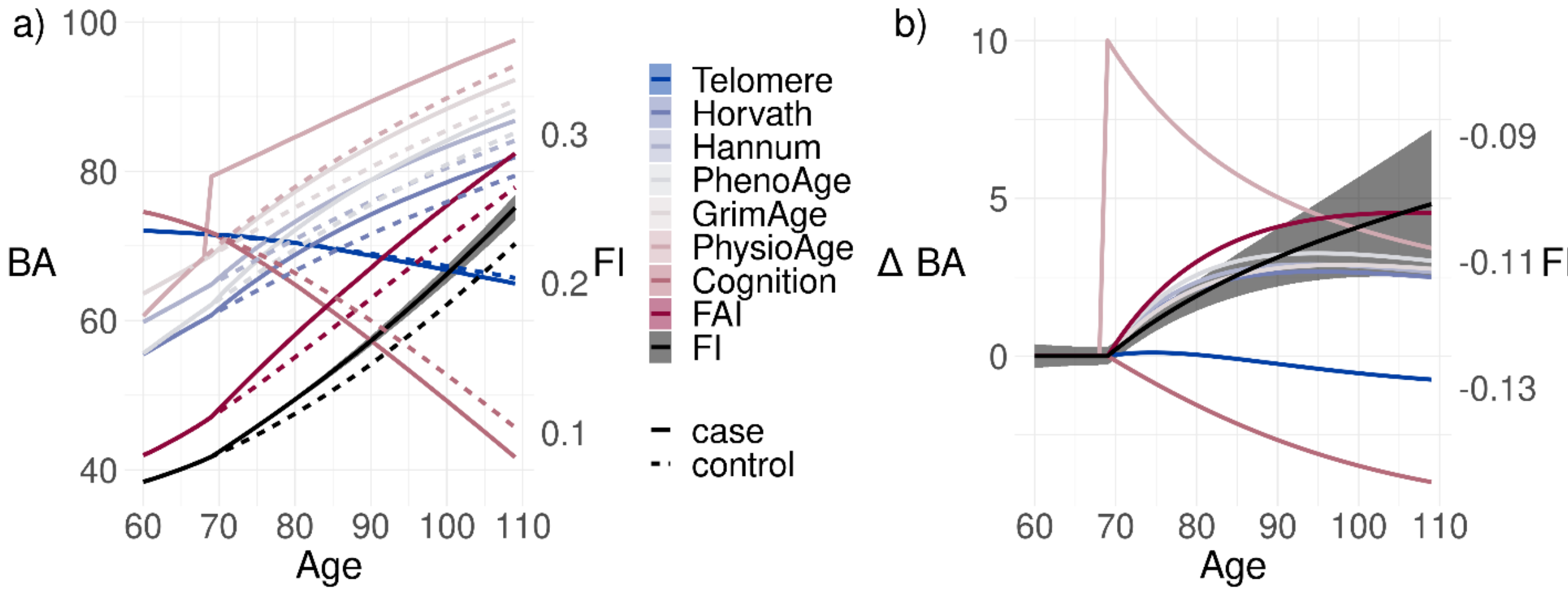

Figure S15: **Simulated harmful intervention on PhysioAge.** We simulated a hypothetical intervention at age 70 which instantly ages PhysioAge by 10 years. The effects are identical to the rejuvenation (Figure 4) but with the sign flipped. Step size: 1 year. Band is standard error (often smaller than line width). The FI has its own y-scale as indicated on the right-hand side.

Both the rejuvenation and worsening of PhysioAge show delayed and transient effects in the other BAs due to the information of the intervention propagating from PhysioAge throughout the network. The natural variables, $z_k$, have very different dynamics. This is because they are eigenvectors ("normal modes") which do not share information (in the mean). While multiple $z_k$ can receive correlated information through the noise, that information averages to 0. As a result, intervening on any single $z_j$ leaves the other $z_k$ permanently unaffected, Figure S16b. All BAs driven by $z_1$ via $\vec{b} = \boldsymbol{P}\vec{z}$ are commensurately affected by the intervention, Figure S16a. Because $z_1$ is well-connected, it drives *all* of the BAs when intervened upon. This means that the interventions on $z_1$ appear optimal, but can look complicated in the BAs — only when monitored in the z-picture do they appear simple.

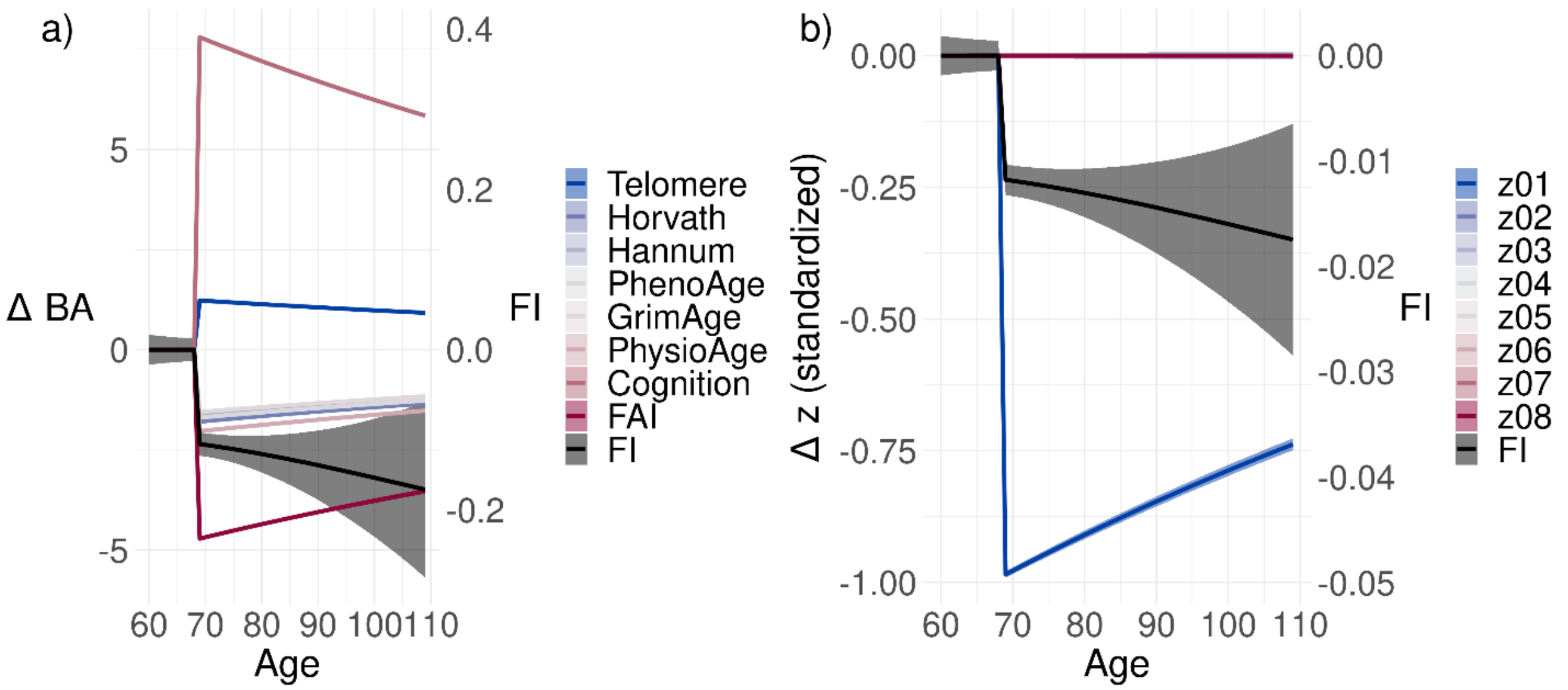

Figure S16**: Simulated intervention on the least stable natural variable, $\boldsymbol{z_1}$**. We simulated a hypothetical intervention at age 70 which instantly rejuvenated $z_1$ by 10 years. The $z_j$ are dynamically independent hence the intervention never affects the $z_{j\neq 1}$, greatly simplifying its effects. The intervention is equivalent to intervening on all of the BAs with weighs equal to $P_{.1}$ (because $\vec{b} = \boldsymbol{P}\vec{z}$). The FI continues to improve after intervention, since it is unstable and the initial change compounds with age. Step size: 1 year. Band is standard error (often smaller than line width). The FI has its own y-scale as indicated on the right-hand side.

For those interested in simulating their own interventions, the GitHub page includes parameters in CSV files and the code needed for simulating interventions. For simulation starting values, we sampled real values[9], but we alternatively provide multivariate normal starting statistics in the repository.

**Sensitivity Analysis**

We tested the robustness of our key results, i.e. the network topology, by fitting with modified data. We considered splitting the males and females, and also we considered adding the transformed FI as a BA and including CA (chronological age).

**Sex-specific Networks**

We considered the confounding effect of sex by separately fitting to males and females. Our key results were unchanged: the networks were similar to the pooled fit and to each other (Figure S17) and showed nearly identical eigenvalues (Figure S18). In particular, both males and females showed 2 low-stability eigenvalues. Males and females also showed high-connectivity of PhysioAge, and to a lesser extent GrimAge. The only noteworthy difference was the strength of the outgoing links from PhysioAge, which was higher in males. When building PhysioAge, males and females were separately calculated using different covariates, following feature selection[9]. Hence these differences may simply reflect that a different definition for PhysioAge was used for males versus females.

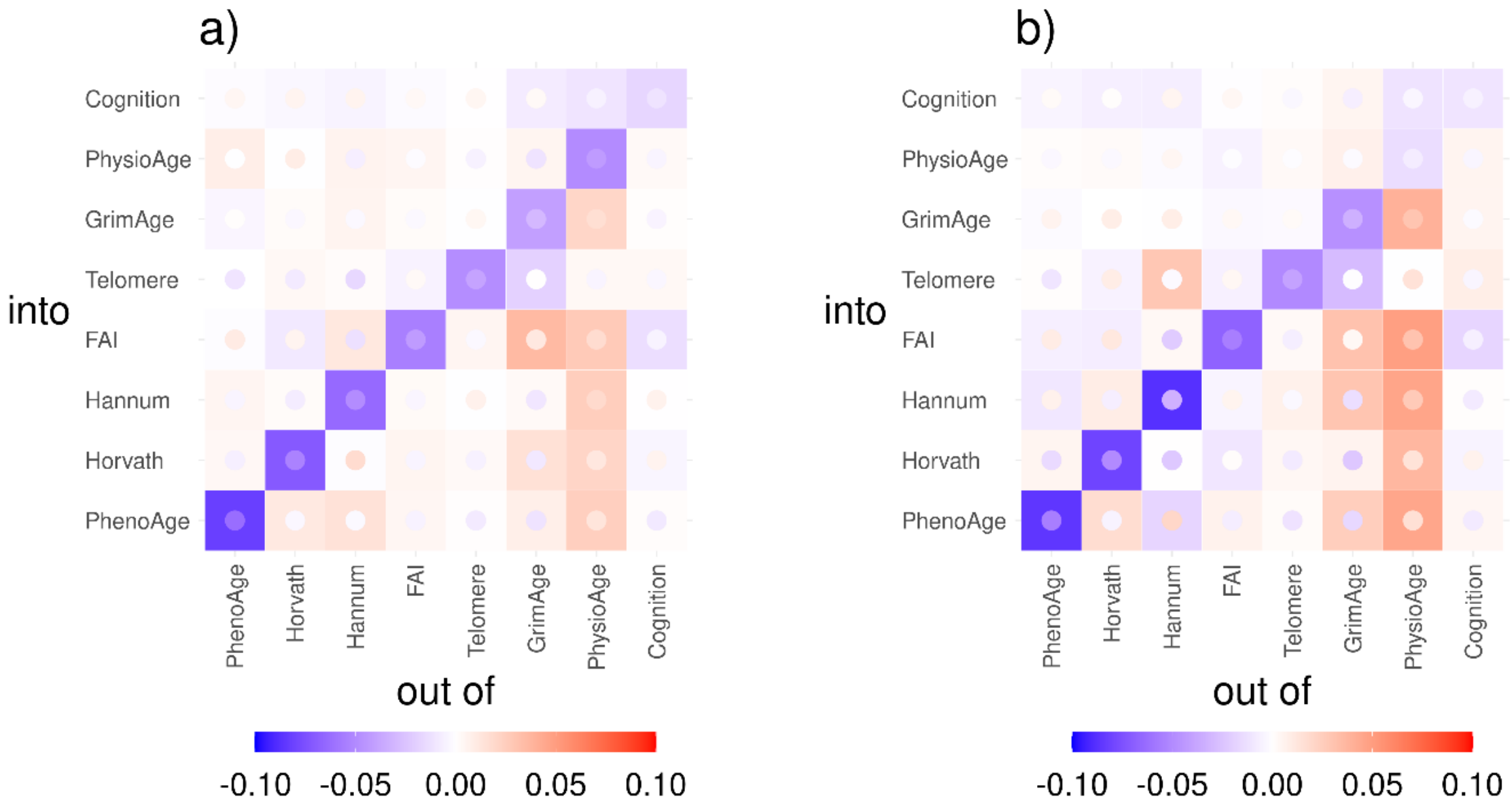

Figure S17: **Female vs male interaction networks**. While females **(a)** and males **(b)** share common network structures, there were notable differences. In particular, while PhysioAge was the main driver in both sexes, males had notably stronger connections from PhysioAge to the BAs. This may reflect that different biomarkers were used to construct the male and female PhysioAges. Inner point is limit of 95% CI closest to 0: point is most visible for the least significant tiles; if point is opposite colour to tile then element is not significant.

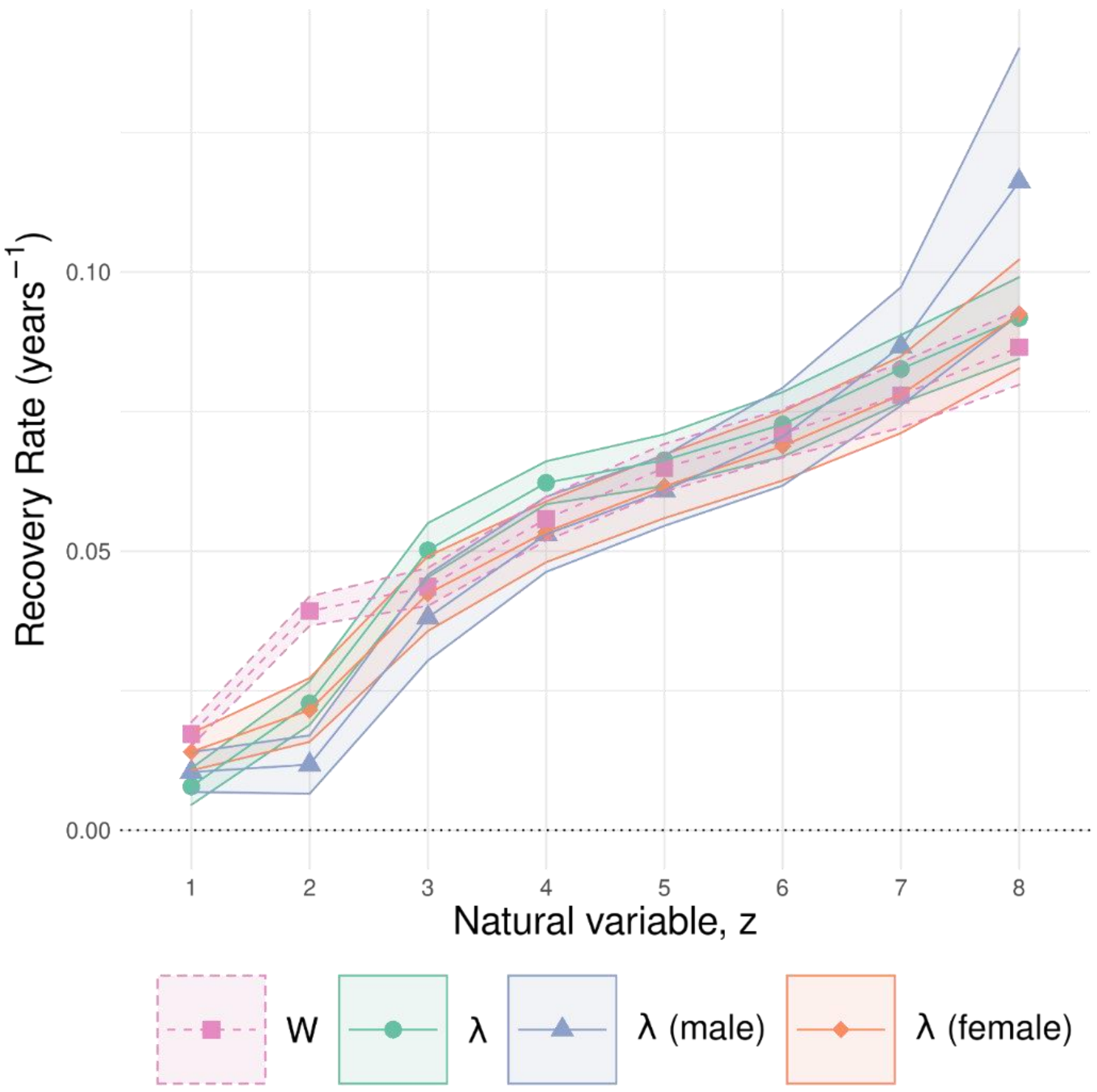

Figure S18: **Network stability (resilience) including sex.** We observe 2 weakly stable eigenvalues which are notably less stable than the least stable network diagonal elements ($W_{ii}$). No strong sex effects are apparent. Bands are standard error.

**Age-specific Networks**

Do the $\boldsymbol{W}$ matrix connections and/or stability change with age? Others have observed age-dependent changes to short timescale resilience[45], or have predicted that long timescale stability will decrease with age[49]. Our estimator is linear and hence does not require a great deal of data to fit with[23]. This permits us to check for age-dependent changes by splitting up the population into age cohorts which are fit separately.

An important consideration is that the missingness was age-dependent, particularly in the epigenetic ages and hence the older ages will be the hardest to estimate and therefore most prone to bias from the imputation. Including dropout, the youngest quartile had 64% missingness including 80% missingness of each epigenetic age versus the oldest quartile which had 87% missingness including 93% missingness of each epigenetic age.

We split up the population into four cohorts based on the baseline age quartiles then fit our model using the usual methodology outline in the main text. In Figure S19 we present the fitted networks using our expectation-maximization imputation strategy from the main text. The MICE imputed data yielded cleaner results, although we are concerned it may have removed the age dependent structure in the data (Figure S20).

The stability is plotted for both the main imputation strategy and MICE in Figure S21. For the main approach we again see an indication that stability may be dropping with age. This change isn't very large relative to errors, however, and is not present in the MICE data.

We see no clear age dependence. We suspect that there is simply too much missing data and not enough individuals/timepoints to confirm or reject any age-dependent changes.

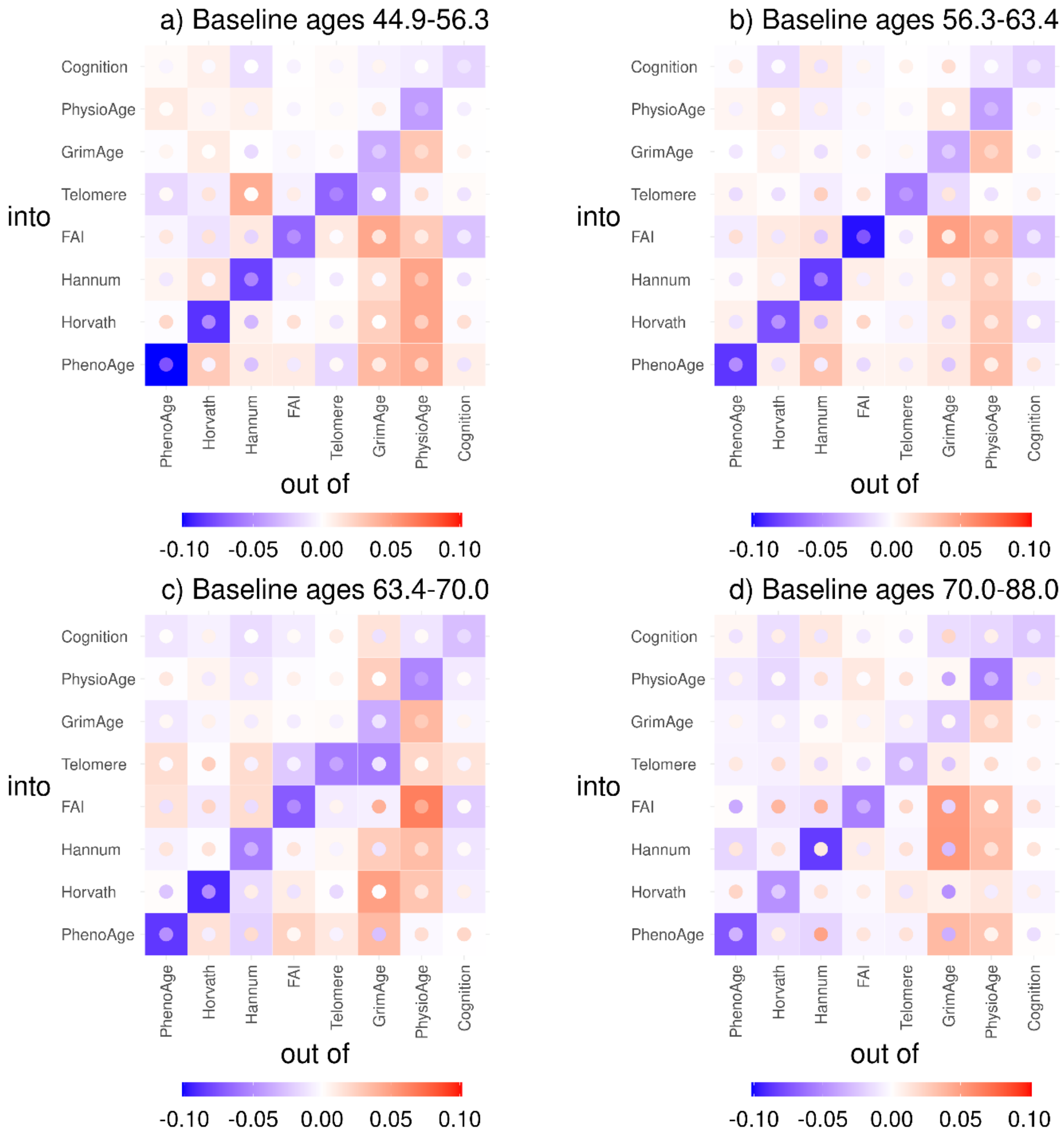

Figure S19: **Network estimates by age-cohort — main result.** We see that across ages, the dominant contributions of outgoing links are from PhysioAge and GrimAge. There are clear differences in sparsity across ages but there is no clear trend. The oldest group, **d**, appears to have a weaker diagonal than the other age ranges. This could be an indication of a loss of resilience or a subtle bias in the imputation model. Inner point is limit of 95% CI closest to 0: point is most visible for the least significant tiles; if point is opposite colour to tile then element is not significant.

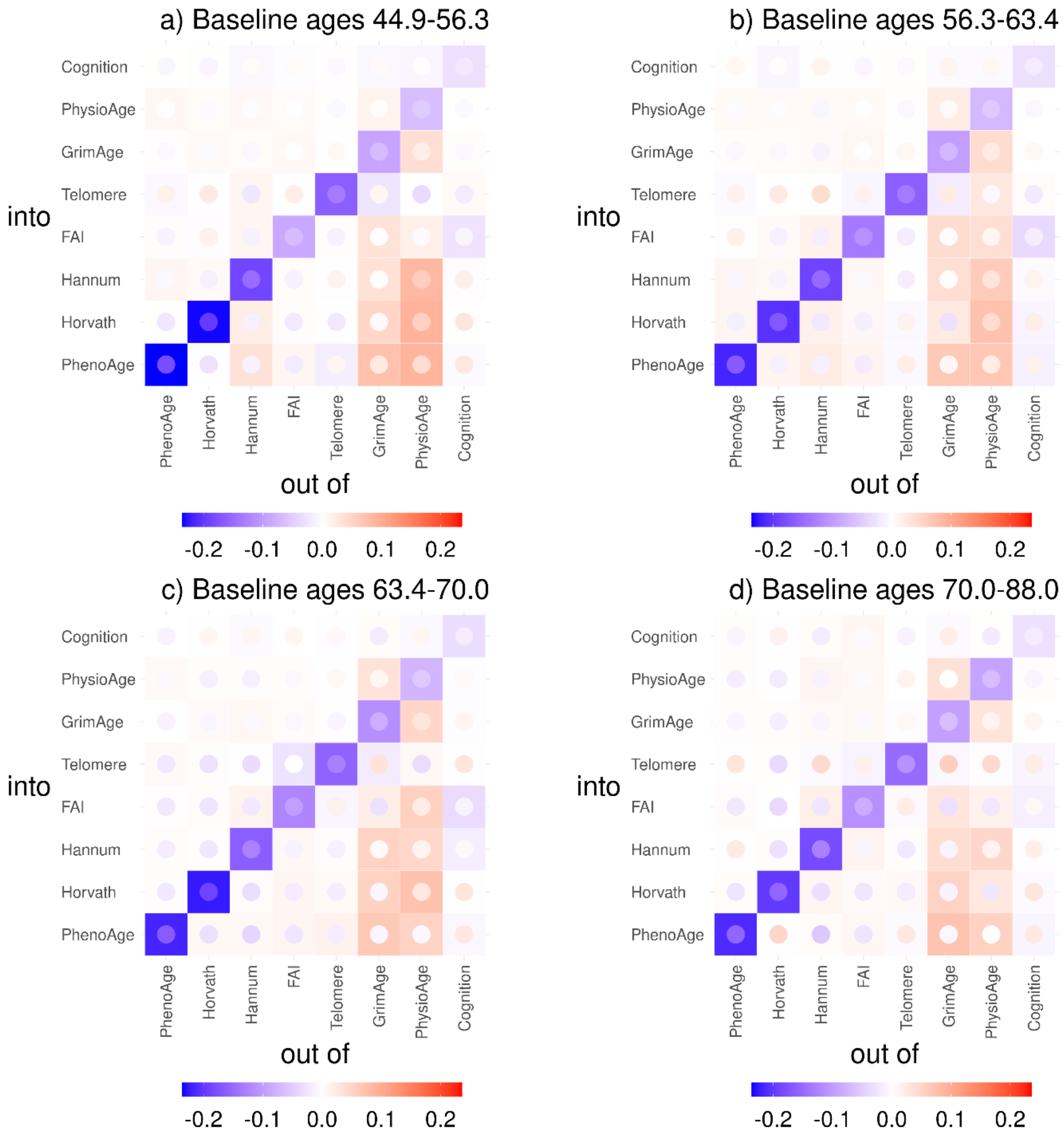

Figure S20: **Network estimates by age-cohort — MICE.** In the present figure we see strong similarity across ages whereas we did not in the main imputation method (Figure S19). Observe that the young network (**a**) coincides well with the main imputation method whereas the older individuals show visually lower agreement. This is an indication that MICE may rely too heavily on young individuals — whom are measured the most often — for training its imputation models, causing it to make older individuals look more like young individuals. This point is discussed in detail in Missing Data. Inner point is limit of 95% CI closest to 0: point is most visible for the least significant tiles; if point is opposite colour to tile then element is not significant.

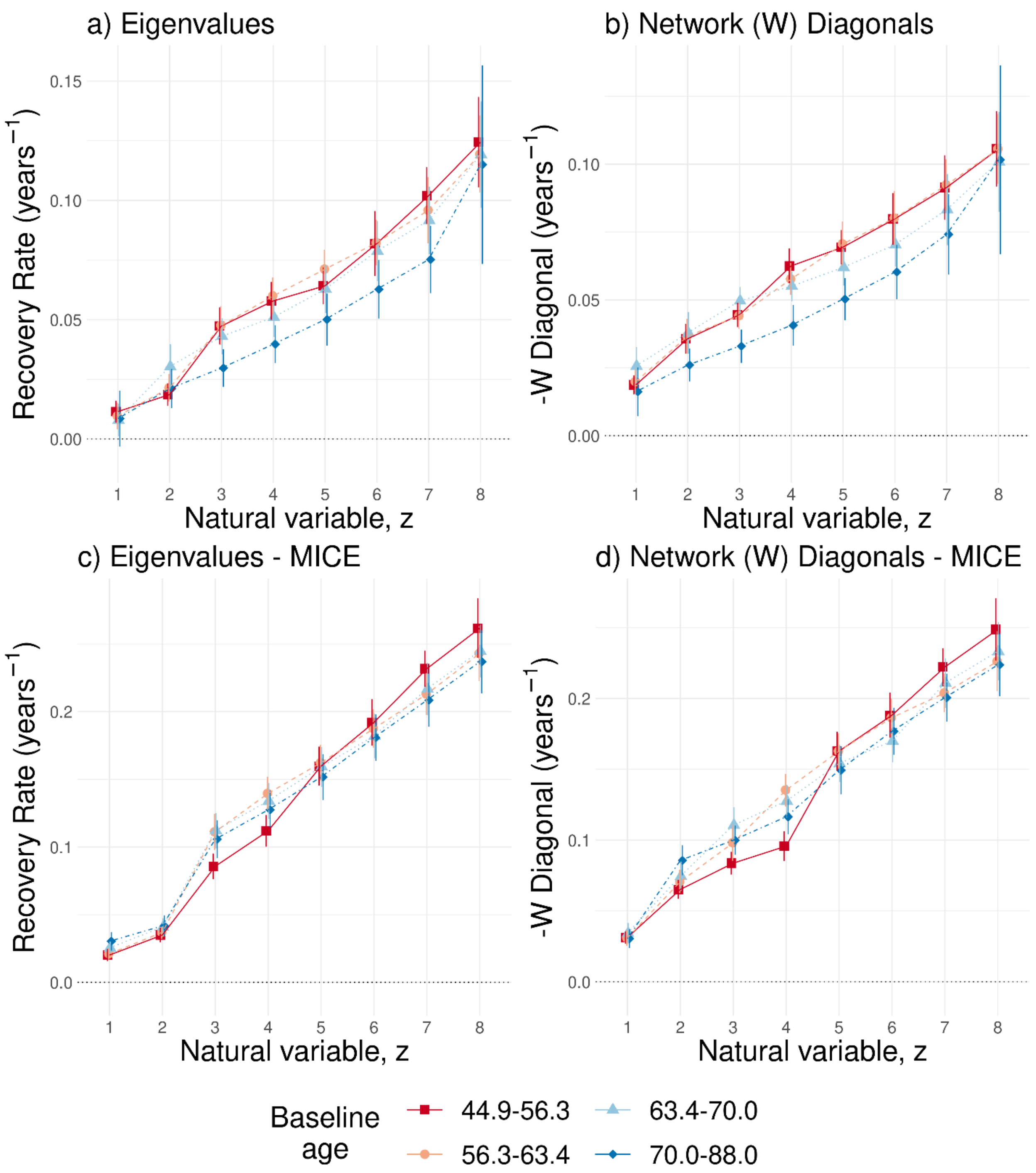

Figure S21: **Age-dependent stability**. Using the main imputation method (**a-b**) or MICE (**c-d**). There may be a trend of increasing or decreasing stability with age but it is typically small relative to the error bars. The most salient feature is the weak diagonal and associated eigenvalues for the oldest group, ages 70-88 in (**a-b**). We are skeptical that this is a real signal and not simply a consequence of the high levels of missingness in this group. In particular, the MICE imputed network does not show any weakening of the diagonal or eigenvalues (**c-d**). If we look at the overall networks we also see that the main imputation method seems to underestimate the diagonal elements of $\boldsymbol{W}$ versus either MICE or available case (Figure S7). Error bars are from bootstrap, 100 resamples; (c-d) also includes error estimate of imputation.

## Chronological Age and the Frailty Index (FI)

The dataset from Li *et al.*[9] includes longitudinal measurements of the FI. In the main analysis we withheld the FI as a proxy for longitudinal changes in health. The FI can be alternatively treated as yet another BA, with the caveat that it should be transformed to prevent issues with fitting (the model assumes normal errors, so it won't fit to FI=0 since it is always assuming symmetric errors above and below the mean). We transformed the FI using the lefthand side of Eq. (S5) with $\gamma/\alpha = 0.065$ and then scaled to match the mean and standard deviations of CA, specifically we multiplied by 22.11 then added 111.2. After transformation, the FI can be treated as normal along with the other BAs. For fitting, we pre-processed the input (BAs and CA) using PCA, picking the optimal number to minimize the 632-RMSE, which was 9 PCs (max: 10). The estimated network is presented in Figure S22. Where we have additionally included CA (age). The main features of the network are unchanged: PhysioAge still plays a central role with a secondary role for GrimAge (i.e. many outgoing links). Age (CA) appears to also have a central role, ostensibly representing unaccounted aging-related degrees-of-freedom. Note that the FI appears to be weakly connected to the others, primarily interacting with FAI. The FAI captures very similar phenomenon to the FI, in particular the FAI is composed of measures which were either directly included in the FI (self-reported hearing and vision) or which modify FI variables (e.g. ADLs)[54,55].

We can use the network to see how information propagates into the FI, a proxy for organism-level health. For example FAI, PhysioAge and Age can directly influence the FI and also gate all incoming information into the FI, whereas e.g. GrimAge must first modify one of those three to affect the FI. The implication is that genetic and epigenetic changes do not directly affect the FI. This is consistent with a recent longitudinal bivariate association study of the FI which found no association with Hannum, Horvath, PhenoAge or GrimAge, although they did find a directed association from DunedinPACE to the FI[37] (DunedinPACE is not included in the present study dataset).

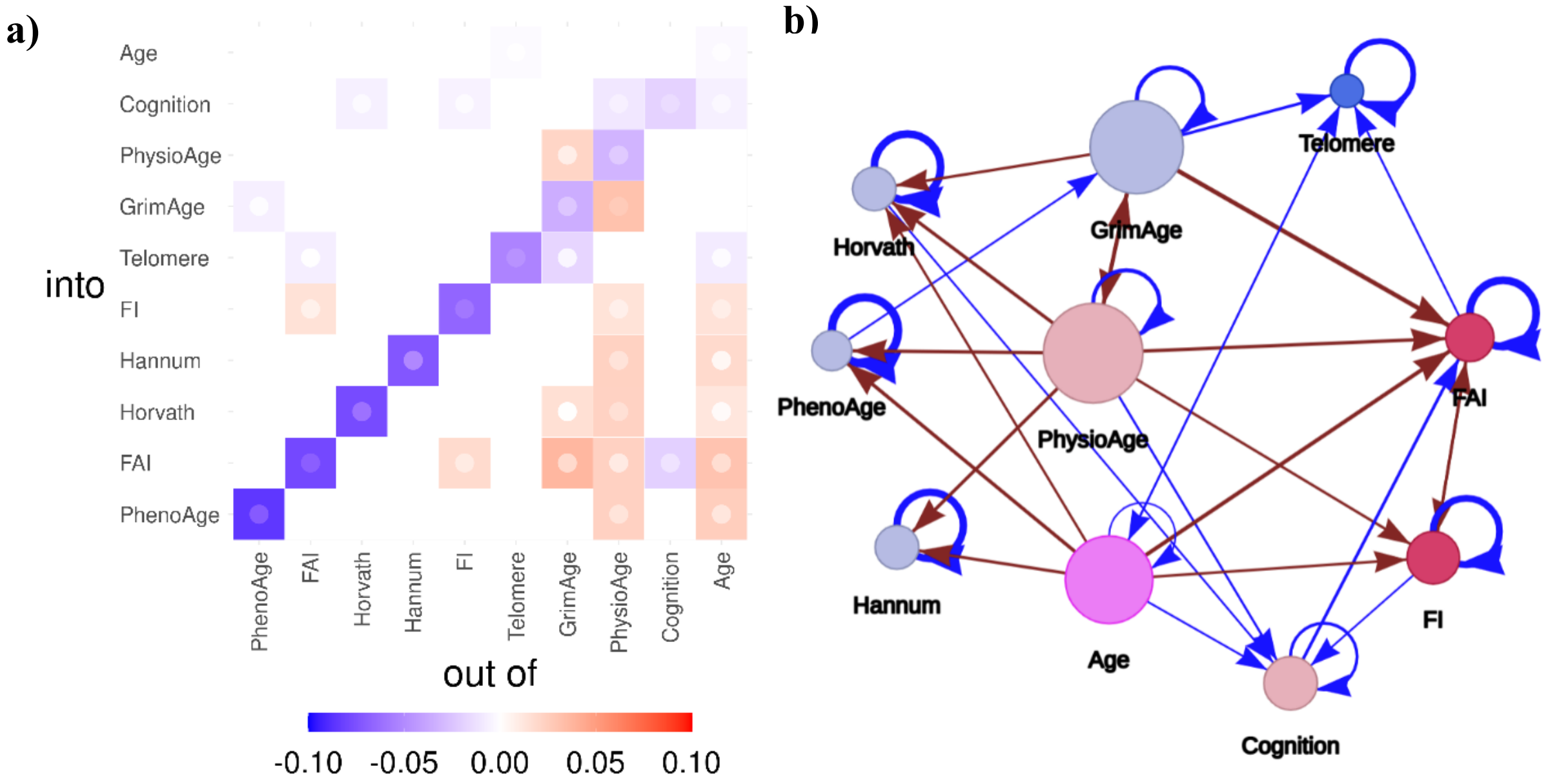

Figure S22: **Network interactome including FI and Age (chronological age)**. **a) Network weight matrix, *W*. b) Network representation.** Age and the FI can be included in the network, allowing us to see how information moves in and out. Observe that the essential features of the network are similar to the main text (Figure 1): with central nodes of PhysioAge and GrimAge. Age is another central node, ostensibly representing unaccounted aging effects that are not captured by the BAs. The FI appears to only connect with higher-level scales: PhysioAge, FAI and Age, having only indirect connections to epigenetic and genetic scales. The FI has been log-scale as described in the text. Both representations **a)** and **b)** are equivalent.

## Works Cited

1. Campisi J, Kapahi P, Lithgow GJ, Melov S, Newman JC, Verdin E. From discoveries in ageing research to therapeutics for healthy ageing. *Nature*. 2019;571(7764):183-192. doi:10.1038/s41586-019-1365-2

2. Zhang B, Trapp A, Kerepesi C, Gladyshev VN. Emerging rejuvenation strategies—Reducing the biological age. *Aging Cell*. Published online December 31, 2021. doi:10.1111/acel.13538

3. Jylhävä J, Pedersen NL, Hägg S. Biological Age Predictors. *EBioMedicine*. 2017;21:29-36. doi:10.1016/j.ebiom.2017.03.046

4. Rutledge J, Oh H, Wyss-Coray T. Measuring biological age using omics data. *Nat Rev Genet*. 2022;23(12):715-727. doi:10.1038/s41576-022-00511-7

5. Levine ME, Lu AT, Quach A, et al. An epigenetic biomarker of aging for lifespan and healthspan. *Aging*. 2018;10(4):573-591. doi:10.18632/aging.101414

6. Lu AT, Quach A, Wilson JG, et al. DNA methylation GrimAge strongly predicts lifespan and healthspan. *Aging*. 2019;11(2):303-327. doi:10.18632/aging.101684

7. Klemera P, Doubal S. A new approach to the concept and computation of biological age. *Mech Ageing Dev*. 2006;127(3):240-248. doi:10.1016/j.mad.2005.10.004

8. Lehallier B, Gate D, Schaum N, et al. Undulating changes in human plasma proteome profiles across the lifespan. *Nat Med*. 2019;25(12):1843-1850. doi:10.1038/s41591-019-0673-2

9. Li X, Ploner A, Wang Y, et al. Longitudinal trajectories, correlations and mortality associations of nine biological ages across 20-years follow-up. *Elife*. 2020;9. doi:10.7554/eLife.51507

10. Jansen R, Han LK, Verhoeven JE, et al. An integrative study of five biological clocks in somatic and mental health. *Elife*. 2021;10. doi:10.7554/eLife.59479

11. Sehgal R, Meer M, Shadyab AH, et al. Systems Age: A single blood methylation test to quantify aging heterogeneity across 11 physiological systems. *bioRxiv*. Published online July 17, 2023:2023.07.13.548904. doi:10.1101/2023.07.13.548904

12. Cohen AA, Ferrucci L, Fülöp T, et al. A complex systems approach to aging biology. *Nature Aging*. 2022;2(7):580-591. doi:10.1038/s43587-022-00252-6

13. Kirkwood TBL. Systems biology of ageing and longevity. *Philos Trans R Soc Lond B Biol Sci*. 2011;366(1561):64-70. doi:10.1098/rstb.2010.0275

14. Kennedy BK, Berger SL, Brunet A, et al. Geroscience: linking aging to chronic disease. *Cell*. 2014;159(4):709-713. doi:10.1016/j.cell.2014.10.039

15. López-Otín C, Blasco MA, Partridge L, Serrano M, Kroemer G. Hallmarks of aging: An expanding universe. *Cell*. 2023;186(2):243-278. doi:10.1016/j.cell.2022.11.001

16. Pierson E, Koh PW, Hashimoto T, et al. Inferring Multidimensional Rates of Aging from Cross-Sectional Data. In: Chaudhuri K, Sugiyama M, eds. *Proceedings of Machine Learning Research*. Vol 89. Proceedings of Machine Learning Research. PMLR; 2019:97-107. http://proceedings.mlr.press/v89/pierson19a.html

17. Farrell S, Mitnitski A, Rockwood K, Rutenberg AD. Interpretable machine learning for high-dimensional trajectories of aging health. *PLoS Comput Biol*. 2022;18(1):e1009746. doi:10.1371/journal.pcbi.1009746

18. Pridham G, Rockwood K, Rutenberg A. Efficient representations of binarized health deficit data: the frailty index and beyond. *Geroscience*. Published online January 27, 2023. doi:10.1007/s11357-022-00723-z

19. Howlett SE, Rutenberg AD, Rockwood K. The degree of frailty as a translational measure of health in aging. *Nature Aging*. 2021;1(8):651-665. doi:10.1038/s43587-021-00099-3

20. Palliyaguru DL, Moats JM, Di Germanio C, Bernier M, de Cabo R. Frailty index as a biomarker of lifespan and healthspan: Focus on pharmacological interventions. *Mech Ageing Dev*. 2019;180:42-48. doi:10.1016/j.mad.2019.03.005

21. Albert R, Jeong H, Barabasi AL. Error and attack tolerance of complex networks. *Nature*. 2000;406(6794):378-382. doi:10.1038/35019019

22. Csete M, Doyle J. Bow ties, metabolism and disease. *Trends Biotechnol*. 2004;22(9):446-450. doi:10.1016/j.tibtech.2004.07.007

23. Pridham G, Rutenberg AD. Network dynamical stability analysis reveals key “mallostatic” natural variables that erode homeostasis and drive age-related decline of health. *Sci Rep*. 2023;13(1):1-12. doi:10.1038/s41598-023-49129-7

24. Ledder G. *Mathematics for the Life Sciences*. Springer New York; 2013. doi:10.1007/978-1-4614-7276-6

25. Ives AR. Measuring Resilience in Stochastic Systems. *Ecol Monogr*. 1995;65(2):217-233. doi:10.2307/2937138

26. Avchaciov K, Antoch MP, Andrianova EL, et al. Unsupervised learning of aging principles from longitudinal data. *Nat Commun*. 2022;13(1):6529. doi:10.1038/s41467-022-34051-9

27. Mitnitski A, Rockwood K. Aging as a process of deficit accumulation: its utility and origin. *Interdiscip Top Gerontol*. 2015;40:85-98. doi:10.1159/000364933

28. Dawid AP. Beware of the DAG! In: *Proceedings of Machine Learning Research*. PMLR; 2010:59-86. http://proceedings.mlr.press/v6/dawid10a.html

29. Yashin AI, Arbeev KG, Akushevich I, Kulminski A, Akushevich L, Ukraintseva SV. Stochastic model for analysis of longitudinal data on aging and mortality. *Math Biosci*. 2007;208(2):538-551. doi:10.1016/j.mbs.2006.11.006

30. Kojima G, Iliffe S, Walters K. Frailty index as a predictor of mortality: a systematic review and meta-analysis. *Age Ageing*. 2018;47(2):193-200. doi:10.1093/ageing/afx162

31. R Core Team. R: A Language and Environment for Statistical Computing. Published online 2021. https://www.R-project.org/

32. Berglund K, Reynolds CA, Ploner A, et al. Longitudinal decline of leukocyte telomere length in old age and the association with sex and genetic risk. *Aging*. 2016;8(7):1398-1415. doi:10.18632/aging.100995

33. Robin X, Turck N, Hainard A, et al. pROC: an open-source package for R and S+ to analyze and compare ROC curves. *BMC Bioinformatics*. 2011;12:77. doi:10.1186/1471-2105-12-77

34. Hanley JA, McNeil BJ. The meaning and use of the area under a receiver operating characteristic (ROC) curve. *Radiology*. 1982;143(1):29-36. doi:10.1148/radiology.143.1.7063747

35. Pridham G, Rockwood K, Rutenberg A. Strategies for handling missing data that improve Frailty Index estimation and predictive power: lessons from the NHANES dataset. *GeroScience*. Published online February 1, 2022. doi:10.1007/s11357-021-00489-w

36. Hardy SE, Allore H, Studenski SA. Missing data: a special challenge in aging research. *J Am Geriatr Soc*. 2009;57(4):722-729. doi:10.1111/j.1532-5415.2008.02168.x

37. Mak JKL, Karlsson IK, Tang B, et al. Temporal dynamics of epigenetic aging and frailty from midlife to old age. *J Gerontol A Biol Sci Med Sci*. Published online October 27, 2023. doi:10.1093/gerona/glad251

38. Mair W, Goymer P, Pletcher SD, Partridge L. Demography of dietary restriction and death in Drosophila. *Science*. 2003;301(5640):1731-1733. doi:10.1126/science.1086016

39. Tobin R, Pridham G, Rutenberg AD. Modelling lifespan reduction in an exogenous damage model of generic disease. *Sci Rep*. 2023;13(1):16304. doi:10.1038/s41598-023-43005-0

40. Sehl ME, Yates FE. Kinetics of human aging: I. Rates of senescence between ages 30 and 70 years in healthy people. *J Gerontol A Biol Sci Med Sci*. 2001;56(5):B198-208. doi:10.1093/gerona/56.5.b198

41. Stolz E, Mayerl H, Hoogendijk EO, Armstrong JJ, Roller-Wirnsberger R, Freidl W. Acceleration of health deficit accumulation in late-life: evidence of terminal decline in frailty index three years before death in the US Health and Retirement Study. *Ann Epidemiol*. 2021;58:156-161. doi:10.1016/j.annepidem.2021.03.008

42. Stolz E, Mayerl H, Muniz-Terrera G, Gill TM. Terminal decline in physical function in older adults. *J Gerontol A Biol Sci Med Sci*. Published online May 6, 2023. doi:10.1093/gerona/glad119

43. Karin O, Agrawal A, Porat Z, Krizhanovsky V, Alon U. Senescent cell turnover slows with age providing an explanation for the Gompertz law. *Nat Commun*. 2019;10(1):5495. doi:10.1038/s41467-019-13192-4

44. Yang Y, Karin O, Mayo A, et al. Damage dynamics and the role of chance in the timing of E. coli cell death. *Nat Commun*. 2023;14(1):2209. doi:10.1038/s41467-023-37930-x

45. Pyrkov TV, Avchaciov K, Tarkhov AE, Menshikov LI, Gudkov AV, Fedichev PO. Longitudinal analysis of blood markers reveals progressive loss of resilience and predicts human lifespan limit. *Nat Commun*. 2021;12(1):2765. doi:10.1038/s41467-021-23014-1

46. Podolskiy D, Molodtcov I, Zenin A, et al. Critical dynamics of gene networks is a mechanism behind ageing and Gompertz law. *arXiv [q-bioMN]*. Published online February 15, 2015. http://arxiv.org/abs/1502.04307

47. Farrell S, Kane AE, Bisset E, Howlett SE, Rutenberg AD. Measurements of damage and repair of binary health attributes in aging mice and humans reveal that robustness and resilience decrease with age, operate over broad timescales, and are affected differently by interventions. *Elife*. 2022;11:e77632. doi:10.7554/eLife.77632

48. Farrell SG, Mitnitski AB, Rockwood K, Rutenberg AD. Network model of human aging: Frailty limits and information measures. *Phys Rev E*. 2016;94(5-1):052409. doi:10.1103/PhysRevE.94.052409

49. Tarkhov AE, Denisov KA, Fedichev PO. Aging clocks, entropy, and the limits of age-reversal. *bioRxiv*. Published online October 11, 2022:2022.02.06.479300. doi:10.1101/2022.02.06.479300

50. Eberhardt F. Introduction to the foundations of causal discovery. *International Journal of Data Science and Analytics*. 2017;3(2):81-91. doi:10.1007/s41060-016-0038-6

51. Pearl J. *Causality*. 2nd ed. Cambridge University Press; 2009. doi:10.1017/CBO9780511803161

52. Olde Rikkert MGM, Dakos V, Buchman TG, et al. Slowing Down of Recovery as Generic Risk Marker for Acute Severity Transitions in Chronic Diseases. *Crit Care Med*. 2016;44(3):601-606. doi:10.1097/CCM.0000000000001564

53. Reynolds CA, Finkel D, McArdle JJ, Gatz M, Berg S, Pedersen NL. Quantitative genetic analysis of latent growth curve models of cognitive abilities in adulthood. *Dev Psychol*. 2005;41(1):3-16. doi:10.1037/0012-1649.41.1.3

54. Finkel D, Sternäng O, Jylhävä J, Bai G, Pedersen NL. Functional Aging Index Complements Frailty in Prediction of Entry Into Care and Mortality. *J Gerontol A Biol Sci Med Sci*. 2019;74(12):1980-1986. doi:10.1093/gerona/glz155

55. Jiang M, Foebel AD, Kuja-Halkola R, et al. Frailty index as a predictor of all-cause and cause-specific mortality in a Swedish population-based cohort. *Aging*. 2017;9(12):2629-2646. doi:10.18632/aging.101352

56. Edemekong PF, Bomgaars DL, Sukumaran S, Levy SB. *Activities of Daily Living*. StatPearls Publishing, Treasure Island (FL); 2021. https://www.ncbi.nlm.nih.gov/books/NBK470404

57. Petersen, Kaare, Brandt and Pedersen, Michael, Syskind. The matrix cookbook. Published online 2012. https://www.math.uwaterloo.ca/~hwolkowi/matrixcookbook.pdf

58. Wood SN. Fast stable restricted maximum likelihood and marginal likelihood estimation of semiparametric generalized linear models. *J R Stat Soc Series B Stat Methodol*. 2011;73(1):3-36. doi:10.1111/j.1467-9868.2010.00749.x

59. White IR, Royston P, Wood AM. Multiple imputation using chained equations: Issues and guidance for practice. *Stat Med*. 2011;30(4):377-399. doi:10.1002/sim.4067

60. van Buuren S. *Flexible Imputation of Missing Data, Second Edition*. CRC Press; 2018. https://play.google.com/store/books/details?id=lzb3DwAAQBAJ

61. van Buuren S, Groothuis-Oudshoorn K. mice: Multivariate imputation by chained equations in R. *J Stat Softw*. 2010;45(3):1-68. https://dspace.library.uu.nl/handle/1874/44635

62. Hastie T, Tibshirani R, Friedman J. *The Elements of Statistical Learning: Data Mining, Inference, and Prediction*. Vol 2nd. Springer; 2017.

63. Cheetham NJ, Penfold R, Giunchiglia V, et al. The effects of COVID-19 on cognitive performance in a community-based cohort: a COVID symptom study biobank prospective cohort study. *eClinicalMedicine*. Published online July 21, 2023. doi:10.1016/j.eclinm.2023.102086